\documentclass[prd,onecolumn,amsmath,amssymb,superscriptaddress,nofootinbib,11pt]{revtex4-1}
\usepackage{url}

\usepackage{epsfig}
\usepackage{amsfonts}
\usepackage{graphicx}
\usepackage{epsfig}
\usepackage{eepic}
\usepackage{amsmath}
\usepackage{amssymb}
\usepackage{color}
\usepackage{bbm}
\usepackage{dcolumn}
\usepackage{bm}
\usepackage{ulem}
\usepackage{mathrsfs}
\usepackage{bbold}
\usepackage{datetime}

\usepackage{overpic} 
\usepackage{rotating}
\usepackage[usenames,dvipsnames]{xcolor}
\usepackage[colorlinks=true,citecolor=Magenta,linkcolor=Green,urlcolor=Green]{hyperref}
\usepackage{lipsum} 
\usepackage{tikz,tikz-3dplot}
\usetikzlibrary{shapes.geometric}
\usepackage{etoolbox} 
\usepackage[capitalize]{cleveref}
\usepackage{extarrows}

\def\bc{\begin{center}}

\def\ec{\end{center}}
\def\be{\begin{eqnarray}}
\def\ee{\end{eqnarray}}
\definecolor{dyellow}{rgb}{1.,0.8,.0}
\definecolor{myblue}{rgb}{.1,.1,.7}
\definecolor{dcyan}{rgb}{.0,.6,.6}
\definecolor{dmagenta}{rgb}{0.6,0.0,0.6}
\definecolor{brown}{rgb}{0.6,0.2,0.}
\definecolor{darkblue}{rgb}{.0,.0,0.5}
\definecolor{darkred}{rgb}{0.75,0.0,0.0}
\definecolor{orange}{rgb}{1.,.6,.0}
\definecolor{dorange}{rgb}{0.8,.4,.0}
\definecolor{darkgreen}{rgb}{0.0,0.6,0.0}
\definecolor{purple}{rgb}{.4,.0,.4}
\definecolor{lightgrey}{rgb}{0.7, 0.7, 0.7}
\definecolor{grey}{rgb}{0.4, 0.4, 0.4}

\usepackage{geometry}
\usepackage[position=t, singlelinecheck=off]{subfig}
\usepackage[font=small,labelfont=bf,justification=raggedright]{caption}

\newcommand{\xdownarrow}[1]{%
  {\left\downarrow\vbox to #1{}\right.\kern-\nulldelimiterspace}
}
\newcommand{\xuparrow}[1]{%
  {\left\uparrow\vbox to #1{}\right.\kern-\nulldelimiterspace}
}

\begin{document}
\newsavebox{\lefttempbox}
\title{Holographic Superfluid Ring with a Weak Link}
\author{Zhi-Hong \surname{Li}} \email{lizhihong@buaa.edu.cn}
\author{Huai-Fan Li} \email{huaifan999@163.com}
\affiliation{Department of Physics, Shanxi Datong University, Datong 037009, China}
\affiliation{Institute of Theoretical Physics, Shanxi Datong University, Datong, 037009, China}
\begin{abstract}
{\centering {\bf Abstract}\\}
We explore the generation of topological defects in the course of a dynamical phase transition in a ring with a weak link, i.e., a SSS Josephson junction, from the AdS/CFT correspondence. By setting different parameters of the junction (width, steepness, depth) and the final temperature of the quench, the configurations of the charge density and condensate of the order parameters of the dual field theory are presented. Meanwhile, we observe that in the final equilibrium state, variations in parameters of the junctions only affect the configurations of the charge density and condensate of the order parameters, without altering their values outside the junction. However, variations in the final temperature will directly affect the values of the charge density and condensate of the order parameters outside of the junction. Moreover, in the final equilibrium state, we propose an analytic relation between the gauge-invariant velocity in the  two superconducting states in the SSS Josephson junction, which agrees well with the numerical results.


\end{abstract}

\maketitle
\section{introduction}
A weak link dividing two superconductors makes up a Josephson junction \cite{Josephson:1962}. Such a weak link can be, for example, an insulating barrier (corresponding to the superconductor-insulator-superconductor (SIS) junction), a normal conductor (corresponding to the superconductor-normal-superconductor (SNS) junction), or even a very narrow superconductor (corresponding to the SSS junction). Assisted by the AdS/CFT correspondence \cite{Maldacena:1997re,Witten:1998qj,Gubser:1998bc}, which is a useful technique that allows us to solve the strongly coupled field theories from one higher dimension of classical gravity, the holographic Josephson junction was first proposed in \cite{Horowitz:2011dz}, in which they identified a conventional relation between the tuning current and the phase difference of the condensation across the junction after building a (2+1)-dimensional holographic model of a Josephson junction. It was then extensively expanded to a number of models, a 4-dimensional Josephson junction that was discovered in \cite{Wang:2011rva,Siani:2011uj}; a holographic p-wave Josephson junction in \cite{Wang:2011ri} has been studied; an investigation was conducted to examine a holographic model of a (1+1)-dimensional SIS Josephson junction in \cite{Wang:2012yj}; a model of a non-relativistic holographic Josephson junction was built using a Lifshitz geometry in \cite{Li:2014xia}; and other interesting models can be found in \cite{Kiritsis:2011zq,Domokos:2012rj,Cai:2013sua,Liu:2015zca}. However, these works were done in static cases.

In this paper, we will explore the spatial 1D SSS Josephson junction dynamically. 
In particular the winding numbers of the order parameter will turn out dynamically and stochastically in the final equilibrium state.
The Kibble-Zurek mechanism (KZM) \cite{Kibble:1976sj,Kibble:1980mv,Zurek:1985qw} provides an intuitive means to realize this dynamic process. It argues that as the system is approaching the critical temperature $T_c$ from above, its dynamics almost freezes as soon as it goes to the critical slowing down regime, and topological defects emerge in the symmetry-breaking phase.
 KZM has already been extensively tested in a wide range of systems \cite{Chuang:1991zz,Ruutu:1995qz,Carmi:2000zz} and multiple numerical investigations \cite{Laguna:1996pv,Yates:1998kx,Donaire:2004gp}. Holographic investigations on KZM in spatial one-dimensional and two-dimensional systems were specifically conducted in \cite{Sonner:2014tca,Chesler:2014gya,Zeng:2019yhi,Li:2019oyz,Xia:2020cjl,delCampo:2021rak,Li:2021iph,Li:2021dwp,Li:2021mtd,Xia:2021xap,Li:2021jqk,delCampo:2022lqd,Li:2022tab}. For instance, the holographic KZM in a spatial 1D ring was examined by the authors in \cite{Sonner:2014tca}. The holographic KZM for vortices in a superfluid was examined by the authors in \cite{Chesler:2014gya}. A $(2+1)$-dimensional holographic superconductor was analyzed by the authors in \cite{Zeng:2019yhi} for breaking of $U(1)$ symmetry, and they discovered the emergence of topological defects—fluxons with quantized fluxes trapped inside the vortices. Further research on KZM in holographic superconductor and superfluids rings can be seen in \cite{Li:2021mtd,Li:2021jqk,Li:2022tab}. Reviews are available at \cite{kibblereview,zurekreview}.

The aim of this study is to explore the relationship between the gauge-invariant velocity in two superconducting states of the SSS Josephson junction following a linear temperature quench through a second-order phase transition.
The effects of the different parameters, i.e, the length, depth and steepnes of SSS Josephson junction and the final temperature of the quench on the configurations of the phases of the order parameters and the gauge-invariant velocity at the final equilibrium state will be discussed. Finally, by comparing the relationship between the velocity of the superconducting state $u_1$ and velocity of the narrow superconductor state $u_2$ in the SSS Josephson junction under different parameters,
we have discovered that the velocity between the two phases satisfies the following relationship: $(u_2-u_1) \frac{\mathcal{L}}{2}+\frac{L}{2} u_1=\pi$.

This paper is arranged as below: Section \ref{two} presents the holographic mapping of the superfluid ring with a weak link; Section \ref{three} shows the main numerical results about the relationship between the gauge-invariant velocity in the superconducting state and the narrow superconductor state of the SSS Josephson junction; the conclusions are summarized in Section \ref{four}.

\section{Holographic Mapping}\label{two}
In this paper, we begin with the Abelian-Higgs action \cite{Hartnoll:2008vx}, 
\begin{eqnarray}\label{density}
S=\int d^4x\sqrt{-g} \left( -\frac{1}{4} F_{\mu \nu} F^{\mu \nu} - |D \Psi|^2 - m^2 |\Psi|^2\right).
\end{eqnarray}
Here, the field strength of the U(1) gauge field $A_\mu$ is represented by $F_{\mu\nu}=\partial_\mu A_\nu-\partial_\nu A_\mu$, the complex scalar field is denoted as $\Psi=|\Psi|e^{i\theta}$, and the covariant derivative is given by $D_\mu=\nabla_\mu -iA_\mu$.
To explore the time-dependent behavior of the system, we utilize the Eddington-Finkelstein coordinates within the context of the AdS$_4$ planar black hole \cite{Chesler:2013lia},
\begin{eqnarray}
ds^2 = \frac{1}{z^2} \left(-f(z) dt^2 - 2dtdz + dx^2 +dy^2 \right),
\end{eqnarray}
where $f(z) = 1 - (z/z_h)^3$, with $z$ and $z_h$ representing AdS radial coordinate and the horizon location respectively. We have set the AdS radius $l=1$ and $z_h=1$ for simplicity. The AdS boundary is located at $z = 0$. Therefore, the Hawking temperature can be expressed as $T=3/(4\pi)$. 
We will focus our analysis on the probe approximation, thus, the equations of motion can be expressed as:
\begin{eqnarray}\label{eomofwhole}
D_\mu D^\mu\Psi-m^2\Psi=0, \qquad\nabla_\mu F^{\mu\nu}=i\left(\Psi^* D^\nu\Psi-\Psi{(D^\nu\Psi)^*}\right). 
\end{eqnarray}
Given our interest in a model comprising a one-dimensional ring at the boundary, we impose periodic boundary conditions on all fields along the $x$-direction to emulate the compact nature of the ring, and all fields are homogeneous along $y$-direction. 
Hence, the consistent ansatz for the fields are $\Psi = \Psi(t,z,x), A_{t} = A_{t}(t,z,x), A_x=A_x(t,z,x)$ and $A_z =A_y= 0$. 
The equations take the following explicit forms:
 \begin{eqnarray}
\label{eompsi}
\partial_t \partial_z \psi  - \frac12 \big[( i \partial_z A_t -z-i \partial_x A_x - A_x^2 )\psi  + (f' +2 i A_t )\partial_z \psi  + f \partial_z^2 \psi - 2 i A_x \partial_x \psi+ \partial_x^2 \psi  \big] = 0;~&
\\
\label{eom2}
\partial_t \partial_z A_t - \partial_x( \partial_x A_t  + f \partial_z  A_x  - \partial_t  A_x) 
+ 2 A_t |\psi|^2 + i \big[ f (\psi \partial_z \psi^*-\psi^* \partial_z \psi) -(\psi \partial_t \psi^*-\psi^* \partial_t \psi )\big] = 0;~&
\\
\label{eom3}
\partial_t \partial_z A_x - \frac12 \big[ \partial_z (\partial_x A_t + f \partial_z A_x) + i (\psi \partial_x \psi^*-\psi^* \partial_x \psi ) \big] +  A_x |\psi|^2 = 0;~&
\\
\label{eom1}
\partial_z^2A_t-\partial_z \partial_x A_x  + i (\psi \partial_z \psi^*-\psi^* \partial_z \psi) = 0.~&
\end{eqnarray}
where $\psi=\Psi/z$.
The four equations mentioned above are not independent as they satisfy the subsequent constraint equation:
\begin{eqnarray}
\frac{d}{dt}\text{Eq.\eqref{eom1}}+\frac{d}{dz}\text{Eq.\eqref{eom2}}-2\frac{d}{dx}\text{Eq.\eqref{eom3}}\equiv 2i\left(\text{Eq.\eqref{eompsi}}\times\psi^*-\text{Eq.\eqref{eompsi}}^*\times\psi\right).
\end{eqnarray}
Therefore, there are three independent equations corresponding to three fields $\psi, A_t$ and $A_x$.  Since $\psi=\Psi/z$ is a complex field, this also means that there are four independent real fields for four independent real equations. This further suggests that our selection of the gauge $A_z=A_y=0$ is suitable for configuring the system.

Without loss of generality, we fix the scalar mass squared value to be $m^2= -2$.
The asymptotic behaviors of the scalar fields near $z\to0$ are 
\begin{eqnarray}
\Psi&\sim& \Psi_0 z^{\Delta_-}+\Psi_1 z^{\Delta_+}+\dots, \\
\Delta_{\pm}&=&\frac{1}{2}(3\pm\sqrt{9+4m^2})\nonumber.
\end{eqnarray}
in which $\Delta_{\pm}$ represent the conformal dimension of the dual scalar operator in the boundary field theory.
As $m^2= -2$, we get $\Delta_-=1$ and $\Delta_+=2$. Therefore, the asymptotic behavior of scalar field near $z\to0$ is 
\begin{eqnarray}
\Psi= z\left(\Psi_0+\Psi_1 z+\dots\right),
\end{eqnarray}
From holography, $\Psi_0$ is interpreted as the source of scalar operators on the boundary, while $\Psi_1$ is related to the condensate of the order parameter $\langle O\rangle$. At the boundary $z\to0$, we set $\Psi_0=0$ in the standard quantization \cite{Hartnoll:2008vx} in order to satisfy spontaneous symmetry breaking. For the gauge fields one finds that the asymptotic behaviors near $z\to0$ are
\begin{eqnarray}\label{eomat}
A_\mu&\sim& a_\mu+b_\mu z+\dots
\end{eqnarray}
According to the dictionary of gauge-gravity duality, $a_t$ is interpreted as the chemical potential and $a_x$ is the potentials of the spatial component of gauge fields. Correspondingly, $b_t$ is related to the charge density $\rho$, $b_x$ is the conserved current. At the horizon $z_h$, we set $A_t(z_h)=0$ and ensure regular finite boundary conditions for the other fields. Additionally, we impose Dirichlet boundary conditions such that $A_x=0$ at boundary.

From the holographic superconductor \cite{Hartnoll:2008vx}, in the boundary field theory, increasing the charge density is equivalent to reducing the temperature. By employing dimensional analysis, we find that the black hole temperature $T$ possesses a mass dimension of one, whereas the charge density $\rho$ on the boundary has a mass dimension of two. Consequently, the ratio $T/\sqrt{\rho}$ becomes a dimensionless quantity.
Thus, in order to linearly quench the temperature following the KZM \cite{Kibble:1976sj,Kibble:1980mv,Zurek:1985qw}, with $T(t)/T_c=1-t/\tau_Q$, where $\tau_Q$ denotes the quench rate (or quench strength) , we perform a quench of the charge density $\rho$ as
$\rho(t)={\rho_c}/{\left(1-t/\tau_Q\right)^{2}}$,
where $\rho_c$ represents the critical charge density in the homogeneous and static holographic superconducting system (In this paper, we set $\rho_c=4.06$.).  
Prior to the quench, we thermalize the system thoroughly in order to achieve an initial state in equilibrium. In the thermalization process, we add the Gaussian white noise $\xi(x_i,t)$ into the each field in the bulk with $\langle \xi(x_i,t)\rangle=0$ and $\langle \xi(x_i,t)\xi(x_j,t')\rangle=h\delta(t-t')\delta(x_i-x_j)$, with a small amplitude $h=10^{-3}$. Afterwards, we linearly quench the temperature from $T_i = 1.4T_c$ to various final temperatures $T_f = \{0.6T_c, 0.7T_c, 0.8T_c\}$, the system evolves from a state of normal metal to a superconducting state.
We evolve the system by using the 4th-order Runge-Kutta method with time step $\Delta t=0.1$. 
Along the AdS radial direction $z$, we utilize the Chebyshev pseudo-spectral methods employing 21 grid points. As all fields exhibit periodicity in the $x$-direction, i.e., $x \sim x+L$ where $L$ represents the length of the ring, we employ Fourier decomposition along the $x$-direction using 201 grid points.

\begin{figure}[t]
\centering
\includegraphics[trim=1.cm 5.cm 1cm 5cm, clip=true, scale=0.6]{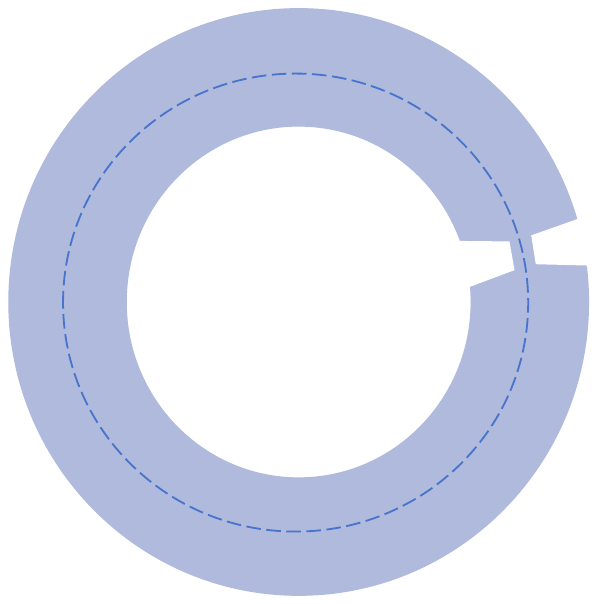}
\put(-360,175){\large Superfluid ring}~
\put(-305,163){$\searrow$}~
\put(-170,103){$\longleftarrow$\large Weak link}~
\caption{Sketchy figure of the superfluid ring with a weak link. }\label{p0}
\end{figure}

In this manuscript, we will dynamically explore the properties of the holographic SSS Josephson junction (refer to Fig.\ref{p0}). To this end, we choose the profile of the charge density $\rho(x)$ as 
\begin{eqnarray}
\rho(x)=\frac{\rho_c}{\left(1-t/\tau_Q\right)^{2}}\left\{1-\frac{1-\epsilon}{2 \tanh(\frac{\mathcal{L}}{2\sigma})}\left[\tanh\left(\frac{x+\frac{\mathcal{L}}{2}}{\sigma}\right)-\tanh\left(\frac{x-\frac{\mathcal{L}}{2}}{\sigma}\right)\right]\right\},
\end{eqnarray}
where $\rho_c$ represents the critical charge density in the homogeneous and static holographic superconducting system, and the parameters \{$\mathcal{L}$, $\sigma$, $\epsilon$\} are the width, steepness and depth of the junction, respectively.

\section{Results }\label{three}
\subsection{Quenched dynamics and winding numbers}
From KZM, it is known that when the system undergoes a quenching process across the critical point and enters the phase of symmetry breaking, it will exhibit the formation of topological defects dynamically and statistically \cite{Kibble:1976sj,Kibble:1980mv,Zurek:1985qw}. Given that our model involves a superfluid ring with a weak link, the topological defect manifests as the winding number of the phase of the order parameter encircling the ring. \cite{Sonner:2014tca, Das:2011cx}.
The definition of the winding number along a compact one-dimensional superfluid ring is as follows:
\be\label{eqw}
W=\oint_\mathcal{C} \frac{d{\bf \theta}}{2\pi} =\oint_\mathcal{C} \frac{\nabla\theta}{2\pi} dx\in\mathbb{Z}
\ee
where \{$\mathcal{C}$, $\theta$\} are the circumference of the ring and the phase of the order parameter respectively. In our numerical calculations, we fix the circumference length at $L=50$ (i.e., $x\in [0, 50]$). According to KZM, while quenching the system across the phase transition point that breaks the $U(1)$ gauge symmetry along a ring, the winding numbers are expected to form \cite{Das:2011cx}. We rapidly quench the system from the initial temperature  to the final temperature, and then keep the system at $T_f$ until it reaches its final equilibrium state.

\begin{figure}[t]
\centering
\includegraphics[trim=1.cm 7.cm 1cm 8cm, clip=true, scale=0.45]{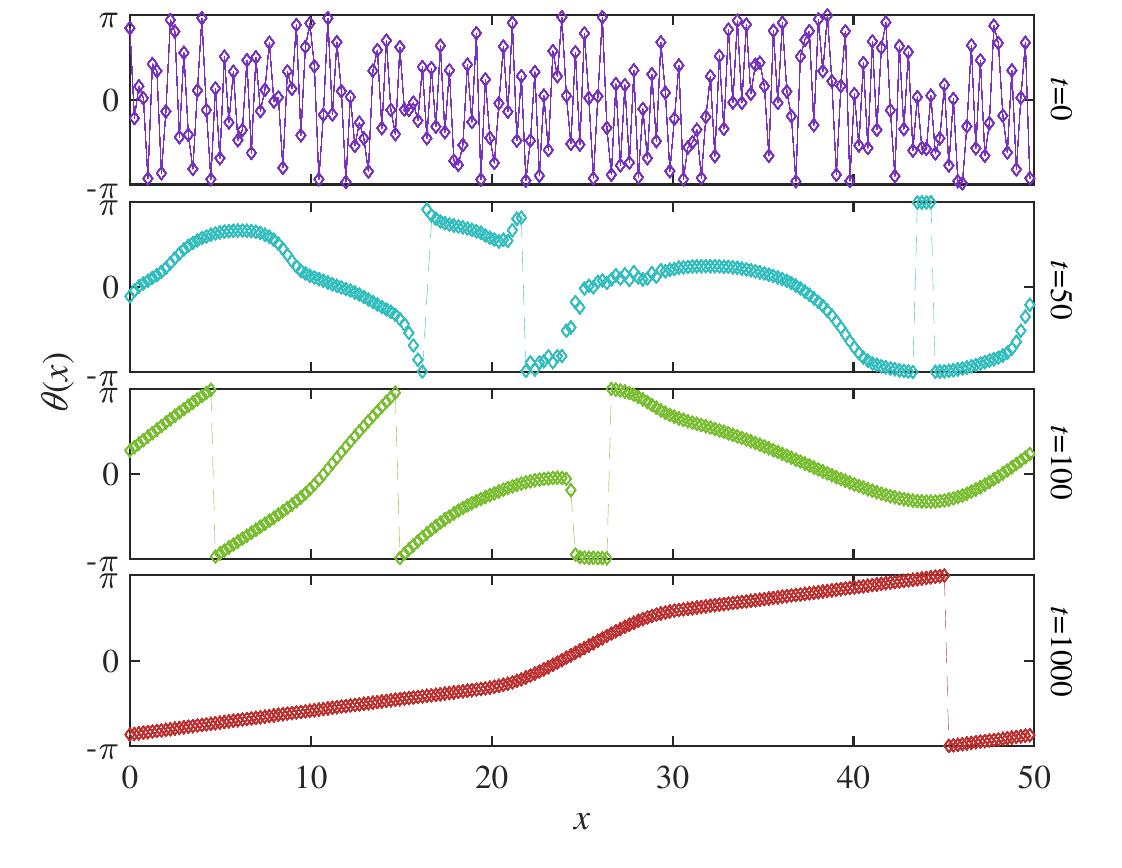}
\put(-220,195){(a)}~
\includegraphics[trim=1.cm 7.cm 1cm 8cm, clip=true, scale=0.45]{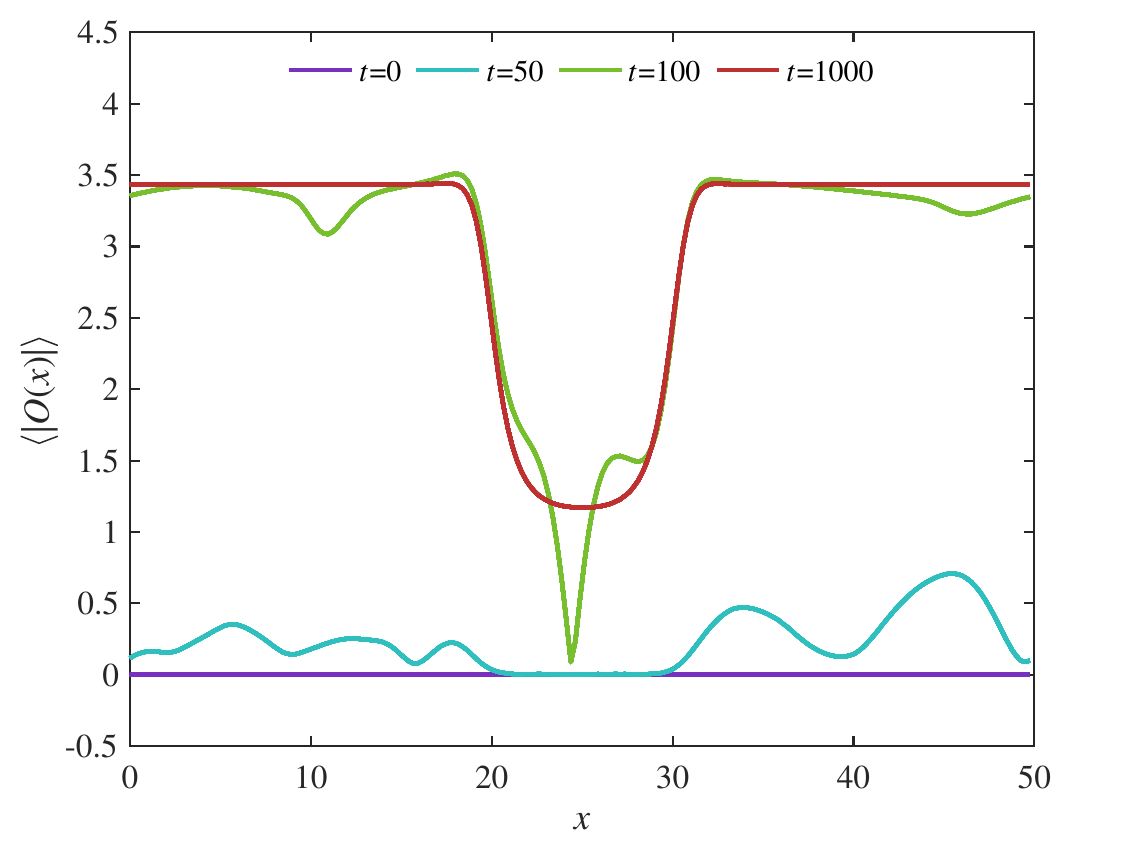}
\put(-220,195){(b)}~
\caption{Temporal evolutions of the phases and the condensate of the order parameter in the holographic superfluid system wrapping around a ring with a weak link. (a) Time evolutions of the phase $\theta$ from the initial time $(t = 0)$ to the final equilibrium state $(t = 1000)$. The phase undergoes a transformation from initial random distributions to ultimately reach a state of equilibrium. The dashed lines represent spurious jumps of the phase at the edges $\theta= \pm\pi$. In the final equilibrium state at time $t=1000$, the winding number is $W = +1$; (b) The growth of the order parameter at four specific times $(t=0, 50, 100, 1000)$. The parameters used for both graphs are \{$\tau_Q=20,\mathcal{L}=10,\sigma=0.5,\epsilon=0.7,T_f=0.8T_c$\} }\label{p1}
\end{figure}

Fig.\ref{p1} shows the temporal evolutions of the phases (panel (a)) and condensate of the order parameter (panel (b)) from the initial time (at temperature $T=1.4T_c$) to the final equilibrium state ($T=0.8T_c$) with the quench rate $\tau_Q=20$.\footnote{The quench rates used in this paper are all $\tau_Q=20$}  The four different colors are corresponding to four specific times ($t=0, 50, 100, 1000$). At the initial time $t=0$, we introduce tiny random seeds of scalar fields into the system and then evolve them while maintaining the temperature at $1.4T_c$ for a period of time, aiming to achieve a state of thermal equilibrium at the initial time. Thus, the phase is randomly distributed in space at $t=0$ and the system is in the normal state with vanishing order parameters. By reducing the temperature below the critical point $T_c$, there will be a spontaneous breaking of the U(1) symmetry, resulting in the emergence of winding numbers along the ring as a consequence of the KZM.

At the time $t=50$ the system has entered a superconducting state, although it remains significantly distant from equilibrium, as evident from panel (b), in which the condensate of the order parameter remain close to 0. The phase $\theta$ at this stage exhibits approximately constant `plateaus', which is a direct consequence of the KZM's prediction that the symmetry will spontaneously break and the phase will randomly choose some constant values in various spatial regions. Since the system is still in the far-from-equilibrium state, the non-equilibrium dynamics may cause the winding numbers to be disrupted or destroyed at this stage for various reasons.

The instant $t=100$ is at the early stage when the condensate of order parameter reaches at the equilibrium value. From  the green line in panel (b) we can see that the absolute value of the condensate of order parameter is close to the value at the time $t = 1000$. Nevertheless, the phase of the order parameter still undergoes dynamical processes until it ultimately reaches a state of equilibrium. For example, at the final equilibrium state $(t=1000)$ the phase becomes `piecewise' smooth lines, contrasting to its appearance at $t=100$, and exhibiting a winding number of $W=+1$.\footnote{We define the winding number as $W=+n$ ($n\geq0$ and $n\in \mathbb{Z}$) when the phase goes from $-\pi$ to $+\pi$ and wraps it $n$ times along the $x$-direction. Conversely, negative winding numbers can be defined in a similar manner. }. 
Due to the presence of the weak link, i.e., Josephson junction in the superfluid ring, the configurations of the phase and condensate of the order parameter will finally be `piecewise' smooth in the final equilibrium state. Therefore, the supercurrent has two constant velocities $\nabla\theta$ along the ring, one is the velocity inside the junction, the other one is outside the junction. The phase is confined within $\theta\in[-\pi, \pi]$, hence, the dashed lines in Fig.\ref{p1}(a) represent the spurious jumps of the phases at the edges $\theta=\pm\pi$. 
\begin{figure}[t]
\centering
\includegraphics[trim=1.cm 7.cm 1cm 8cm, clip=true, scale=0.45]{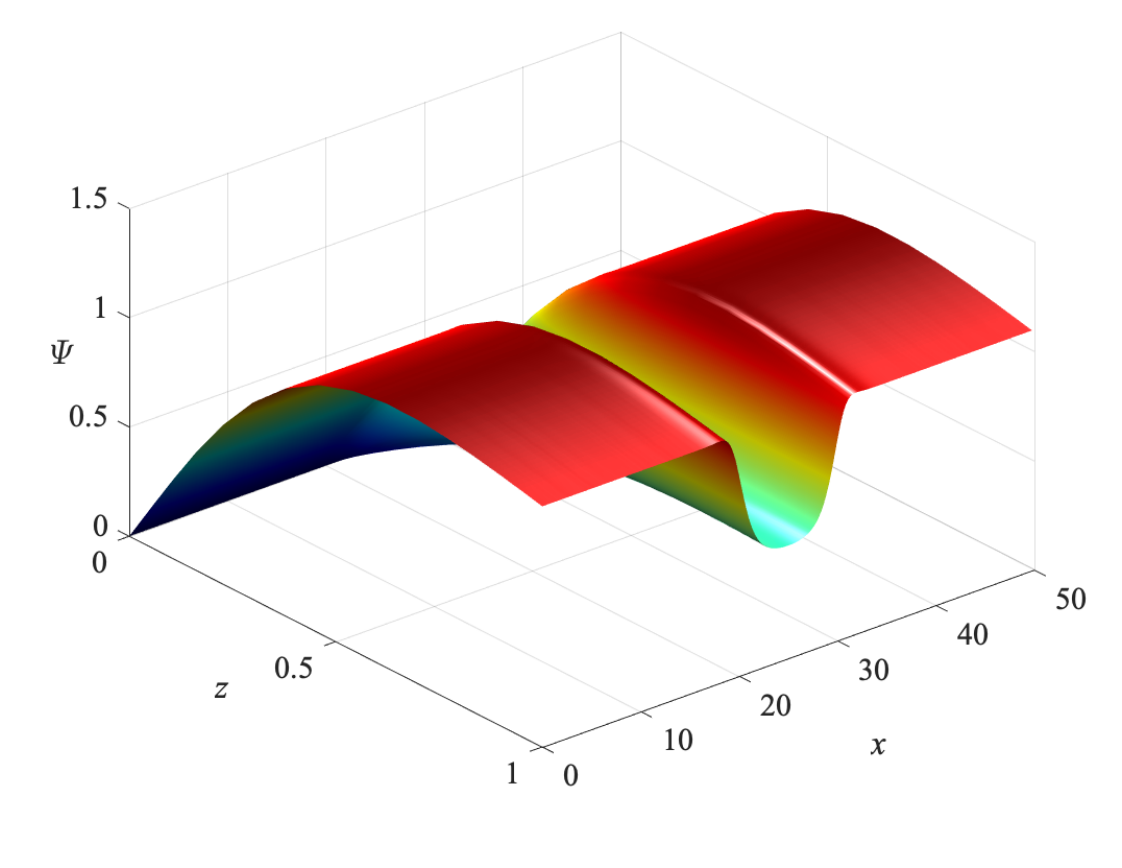}
\put(-220,195){(a)}~
\includegraphics[trim=1.cm 7.cm 1cm 8cm, clip=true, scale=0.45]{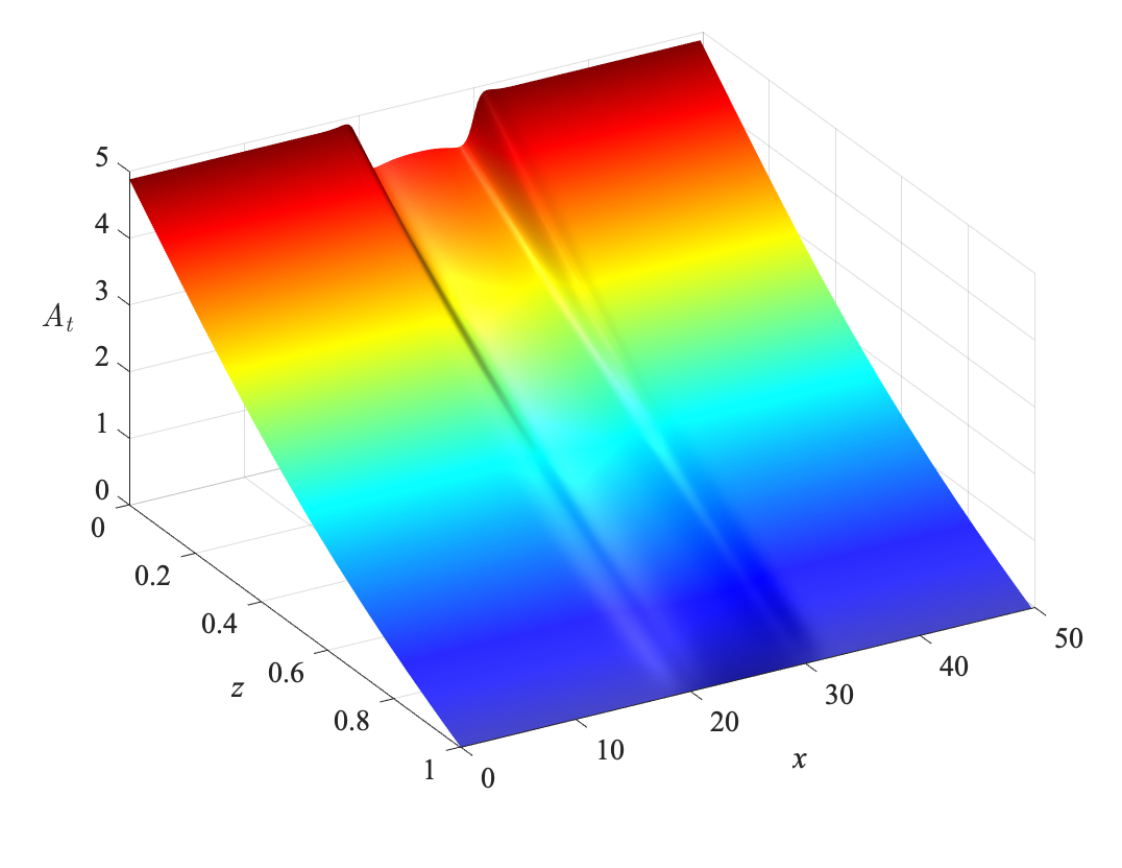}
\put(-220,195){(b)}~
\caption{(a) 3D configurations of $\Psi$ in the AdS bulk at the final equilibrium state. (b) 3D configurations of $A_t$  in the AdS bulk at the final equilibrium state. In both plots we use the parameters \{$\tau_Q=20,\mathcal{L}=10,\sigma=0.5,\epsilon=0.7$ and $T_f=0.8T_c$\}. }\label{p2}
\end{figure}

When rapidly cooling the system through the critical point $T_c$ from high temperature to low temperature, the system will enter a superconducting phase, leading to the emergence of winding numbers as a result of the KZM. In Fig.\ref{p2}(a) we exhibit the  configuration of $\Psi$ as a function of $x$ and $z$ in the AdS bulk at the final equilibrium state. From Fig.\ref{p2}(a), we see that along $x$-direction the order parameter has a `kink' structure in the regime of the junction, while in the radial $z$-direction its behavior is relatively mildly from the horizon $(z = 1)$ to the boundary $(z=0)$. Meanwhile we represent the configuration of $A_t$ as a function of $x$ and $z$ in the AdS bulk at the final equilibrium state in Fig.\ref{p2}(b). 

\subsection{Gauge-invariant velocity }

\begin{figure}[t]
\centering
\includegraphics[trim=1.cm 7.cm 1cm 7cm, clip=true, scale=0.6]{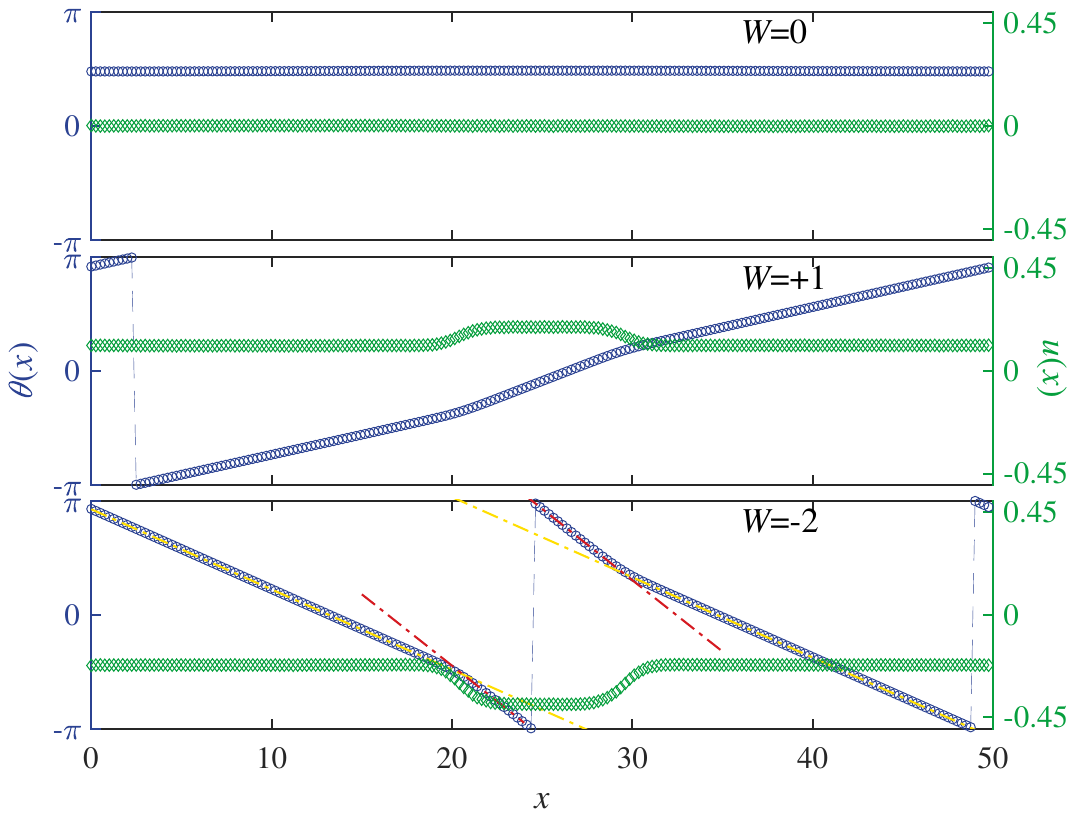}
\caption{Phase configurations $\theta(x)$ (blue circles) and the gauge-invariant velocity $u$ (green diamonds) of the superfluid with various winding numbers ($W = 0, +1$ and $-2$) in the final equilibrium state. In all plots we use $\tau_Q=20$, $\mathcal{L}=10$, $\sigma=0.5$, $\epsilon=0.7$ and $T_f=0.8T_c$. }\label{p3}
\end{figure}

In our model, we impose the Dirichlet boundary conditions for the gauge fields $a_x=0$ at the boundary $z\rightarrow 0$. Therefore, the superfluid velocity precisely corresponds to the gradient of the phase, i.e.,
\be\label{eqwu}
u = \bf\nabla \theta. 
\ee 
In Fig.\ref{p3}, we display  the phase configurations (blue circles) of different winding numbers ($W=0, +1$ and $-2$) of superfluid and their corresponding velocity $u=\nabla\theta$ (green diamonds) at the final equilibrium state with $\tau_Q=20$ . It is clear to see that  the case of $W=0$ is not the same as the other two cases. When $W=0$, both the phase configurations $\theta(x)$  and gauge-invariant velocity $u(x)$ are constant values ($u=0$) at the final equilibrium state, indicating a persistent supercurrent circulating along the ring.  
For the other two cases $W = +1$ and $W = -2$, we can clearly see that the existence of the Josephson junction will lead to the appearance of two different values of superfluid velocity. For instance, in the case of $W=-2$, the yellow dotted dashed line and the red dotted dashed line indicate the gradient of the phase configurations $\theta(x)$ in the superconducting state and the narrow superconducting state inside the junction, respectively. Therefore, the superfluid velocities $u$ are different for the states inside and outside of the Josephson junction for the nonzero winding numbers. 

In this section we focus on the relationship between the gauge-invariant velocity $u(x)$ in the superconducting state and the narrow superconducting state in the SSS Josephson junction by varying different parameters \{$\mathcal{L}$, $\sigma$, $\epsilon$, $T_f$\} at the final equilibrium state. We will focus ourselves on the example of $W = 1$. 
It is obvious from the analysis of the geometry that the gauge-invariant velocity of the superconducting state and the narrow superconducting state in the SSS Josephson junctions satisfy,
\be 
\label{eq:0}
(u_2-u_1) \frac{\mathcal{L}}{2}+\frac{L}{2} u_1=\pi
\ee
where $u_1$ and $u_2$ represent the gauge-invariant velocity in the superconducting state and narrow superconducting state respectively, $\mathcal{L}$ is the width of the junction and $L$ is the length of the ring. Eq.\eqref{eq:0} can be derived from Eq.\eqref{eqw} when $W=1$. Assuming that the phases of inside and outside in the junction are linear, Eq.\eqref{eqw}can be written as 
\be
\oint_\mathcal{C} \nabla\theta dx=\oint_\mathcal{C}  (u_1+u_2) dx=\oint_{(L-\mathcal{L})} u_1 dx+\oint_\mathcal{L} u_2 dx=u_1 (L-\mathcal{L})+u_2 \mathcal{L}=2\pi
\ee
It is then easy to obtain Eq.\eqref{eq:0} by deforming.

{\bf {Case 1: Change width \{$\mathcal{L}=5, 10, 15$\} of the junction }}

\begin{figure}[t]
\centering
\includegraphics[trim=1.cm 7cm 2cm 8cm, clip=true, scale=0.45]{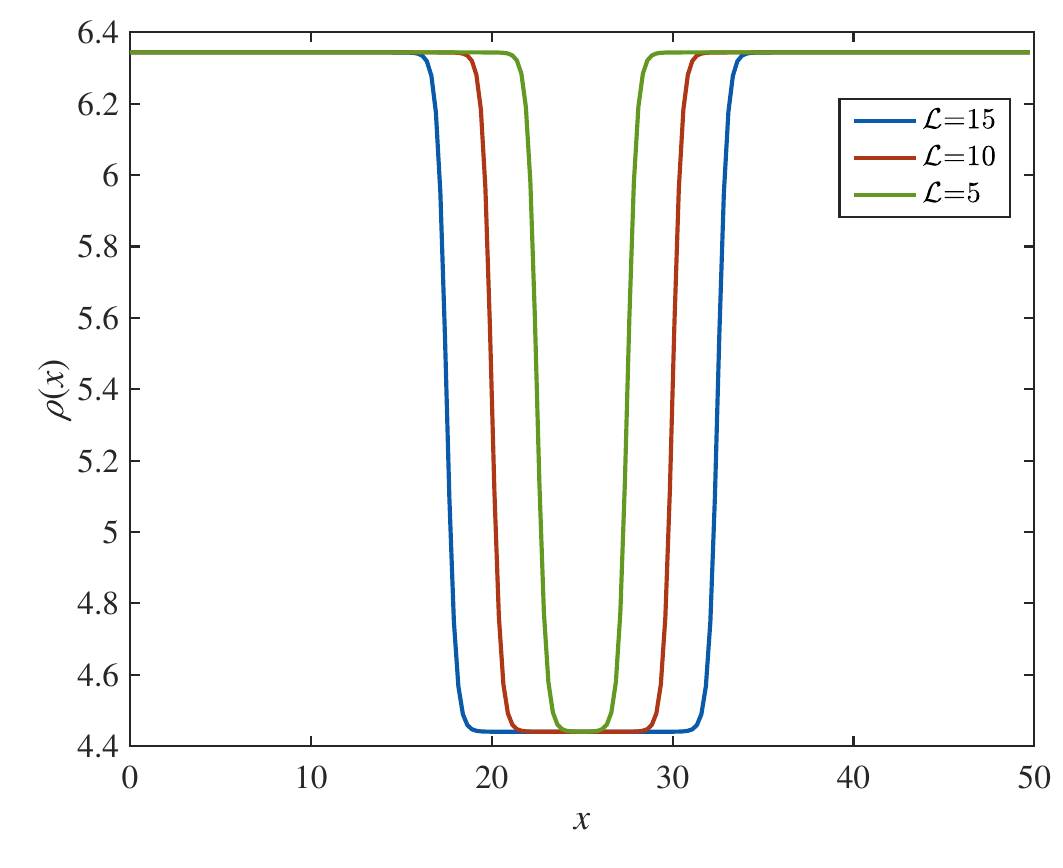}
\put(-220,195){(a)}~
\includegraphics[trim=1.cm 7cm 1cm 8cm, clip=true, scale=0.45]{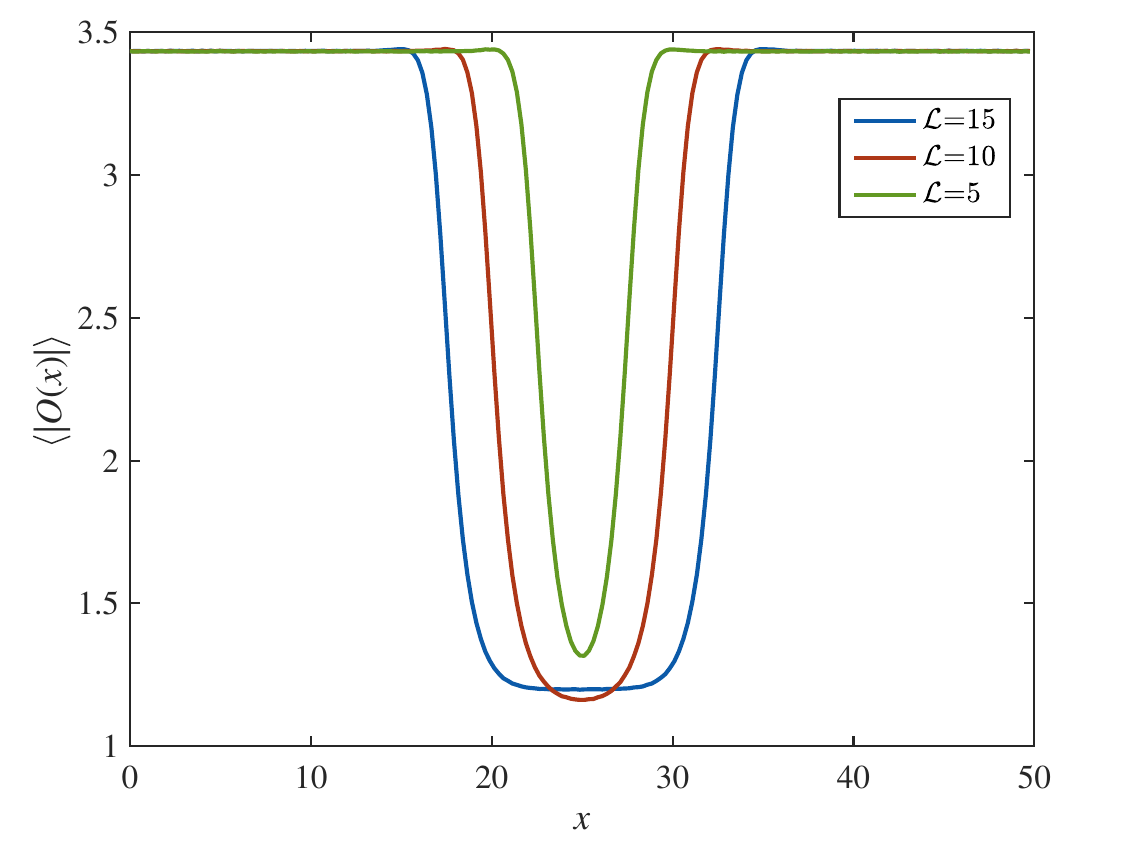}
\put(-220,195){(b)}~ \\
\includegraphics[trim=1.5cm 7cm 0cm 8cm, clip=true, scale=0.3]{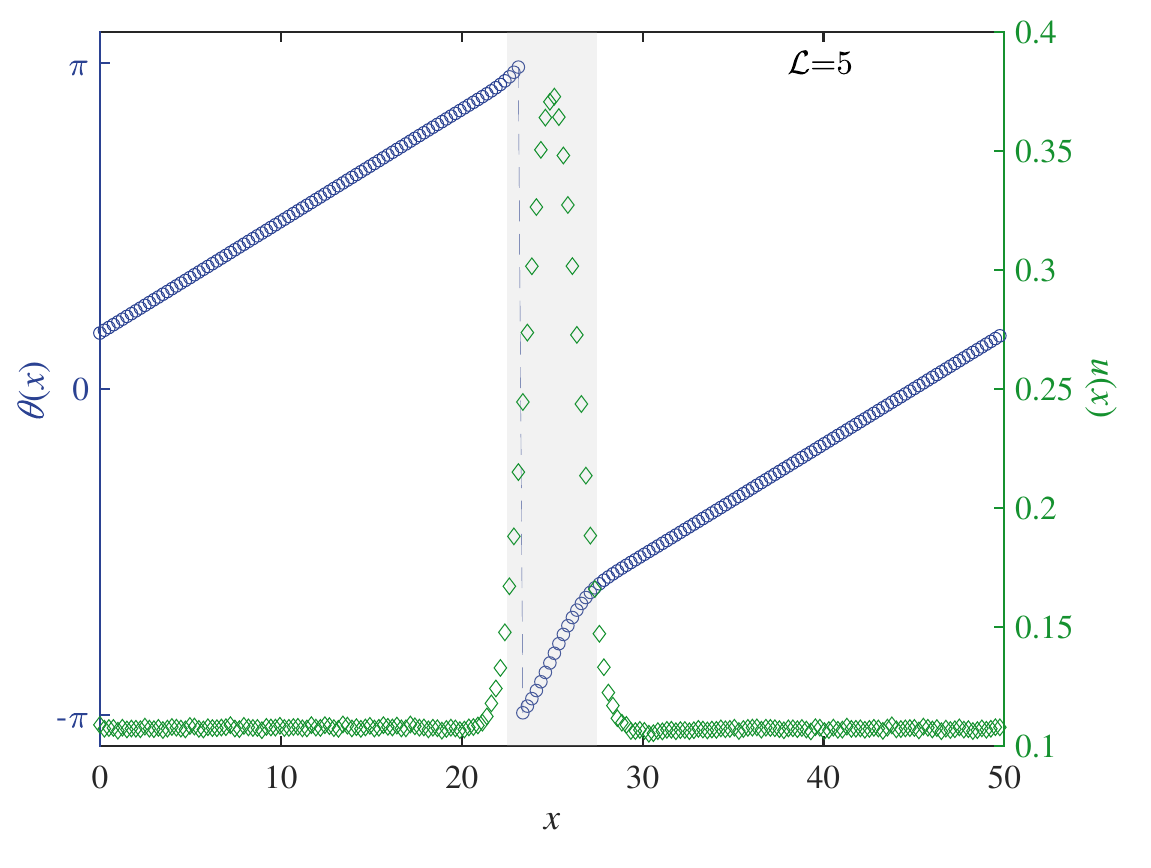}
\put(-150,135){(c)}~
\includegraphics[trim=1.5cm 7cm 0cm 8cm, clip=true, scale=0.3]{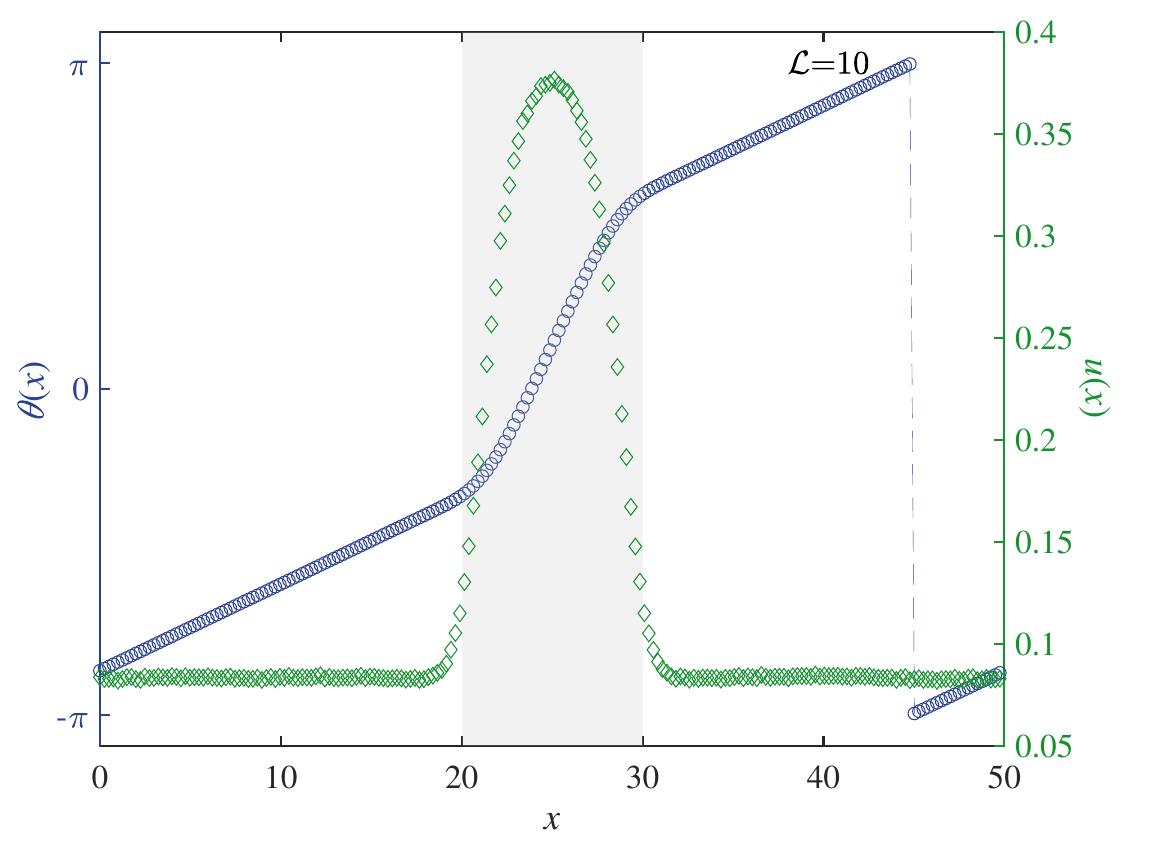}
\put(-150,135){(d)}~
\includegraphics[trim=1.5cm 7cm 0cm 8cm, clip=true, scale=0.3]{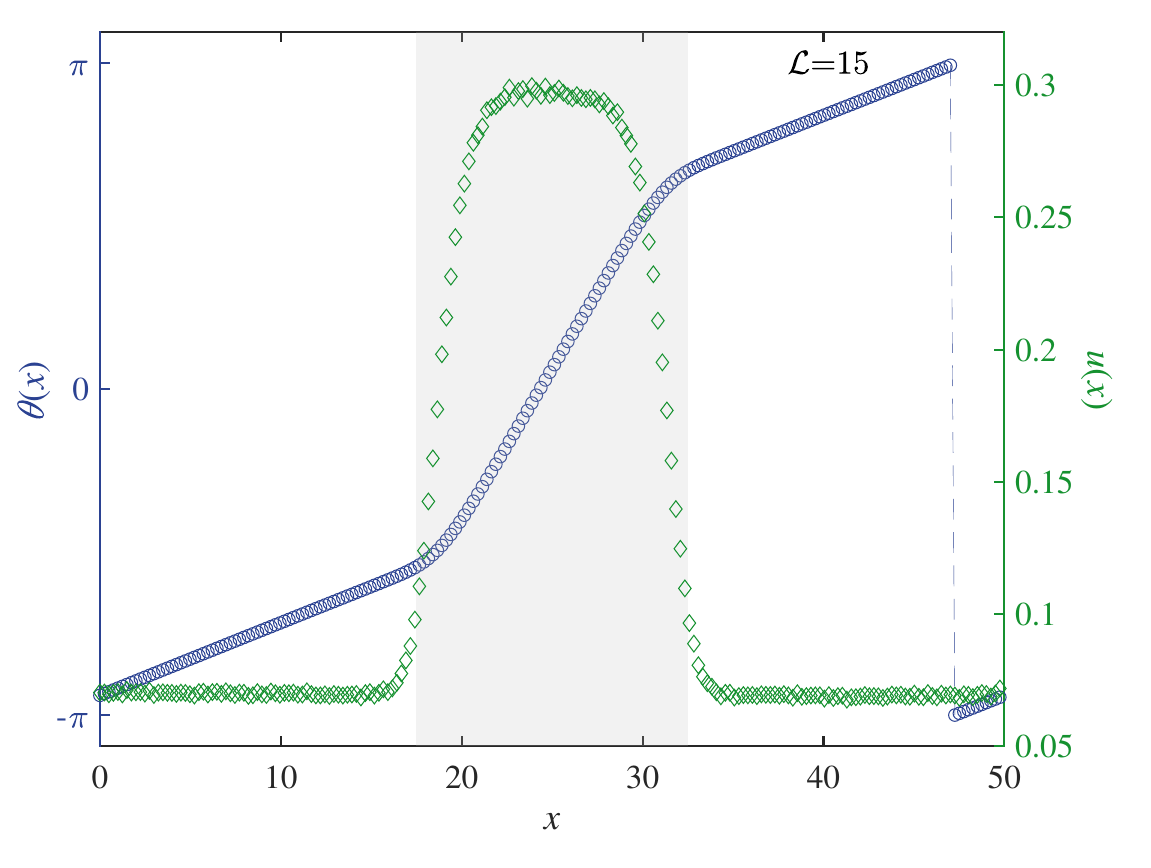}
\put(-150,135){(e)}~
\caption{The charge density $\rho(x)$ (panel (a))  and the condensate of the order parameter $\langle|O(x)|\rangle$ (panel (b)) for different widths $\mathcal{L}=\{5,10,15\}$ in the final equilibrium state; Panel (c)-(e) show the  phase configurations $\theta(x)$(blue circles) and the gauge-invariant velocity $u$ (green diamonds) of the supercurrent with different widths $\mathcal{L}=\{5,10,15\}$ in the final equilibrium state. The shaded areas indicate junction regions. In this figure, other parameters are fixed as  \{$\sigma=0.5$, $\epsilon=0.7$, $T_f=0.8T_c$\}.  }\label{p4}
\end{figure}
Firstly, we study the effect of changing the width of the Josephson junction on the system. 
Fig.\ref{p4} shows the charge density $\rho(x)$ (panel (a)) and the condensate of the order parameter $\langle|O(x)|\rangle$ (panel (b)) for different Josephson junction widths with other parameters fixed as \{$\sigma=0.5$, $\epsilon=0.7$, $T_f=0.8T_c$\}. The green lines, red lines and blue lines represent the different widths \{$\mathcal{L}=5,10,15$\}, respectively. From Fig.\ref{p4}(b) it is clear that the configuration of the condensate of the order parameter $\langle|O(x)|\rangle$ inside the junction broadens as its width increases in the final equilibrium state.
However, the condensate outside of the Josephson junction are nearly indistinguishable because the final temperatures are identical ($T_f=0.8T_c$). Another interesting phenomenon is that for different widths $\mathcal{L}$, the configuration of the phase also changes as the width changes. In Fig.\ref{p4}(c-e) we show the behavior of the phase configurations $\theta(x)$(blue circles) and the gauge-invariant velocity $u$ (green diamonds) for different width \{$\mathcal{L}=5$(panel (c)), $10$(panel (d)), $15$(panel (e))\} with the fixed parameters \{$\sigma=0.5$, $\epsilon=0.7$, $T_f=0.8T_c$\}. Obviously, we can find that the velocities in the superconducting phase ($u_1$) and the narrow superconducting phase ($u_2$) have different values due to the presence of Josephson junctions. For convenience, we provide the approximate values of $u_1$ and $u_2$ for different widths as shown in Fig.\ref{p4} (c)-(e) in table \ref{table1}. From table \ref{table1} we see that the values of the relationship ($(u_2-u_1) \frac{\mathcal{L}}{2}+\frac{L}{2} u_1$) are \{$3.1875 (\mathcal{L}=5)$, $3.1725 (\mathcal{L}=10), 3.1549 (\mathcal{L}=15)$\}, respectively. Their values are close to $\pi$ as Eq.\eqref{eq:0} shows with some errors.

\begin{table}[t]
\renewcommand{\arraystretch}{1.5}
\begin{center}
\begin{tabular}{|c|c|c|c|}
\hline
    $~~~~~~~~~$ & $~~~~u_1~~~~$ & $~~~~u_2~~~~$ & $(u_2-u_1) \frac{\mathcal{L}}{2}+\frac{L}{2} u_1$\\
    \hline
  $\mathcal{L}=5$ & $0.1078$ & $0.3041$ & $3.1857$\\
    \hline
  $\mathcal{L}=10$ & $0.0836$ & $0.3002$ & $3.1725$\\
    \hline
  $\mathcal{L}=15$ & $0.07$ & $0.2573$ & $3.1549$\\
    \hline
\end{tabular}
\end{center}
\caption{Approximate values of $u_1$ and $u_2$ under various widths of the Josephson junction for $\mathcal{L}=(5, 10, 15)$, and other parameters are fixed as  \{$\sigma=0.5$, $\epsilon=0.7$, $T_f=0.8T_c$\}. }
    \label{table1}
\end{table}

\begin{figure}[h]
\centering
\includegraphics[trim=1.cm 7cm 0cm 8cm, clip=true, scale=0.44]{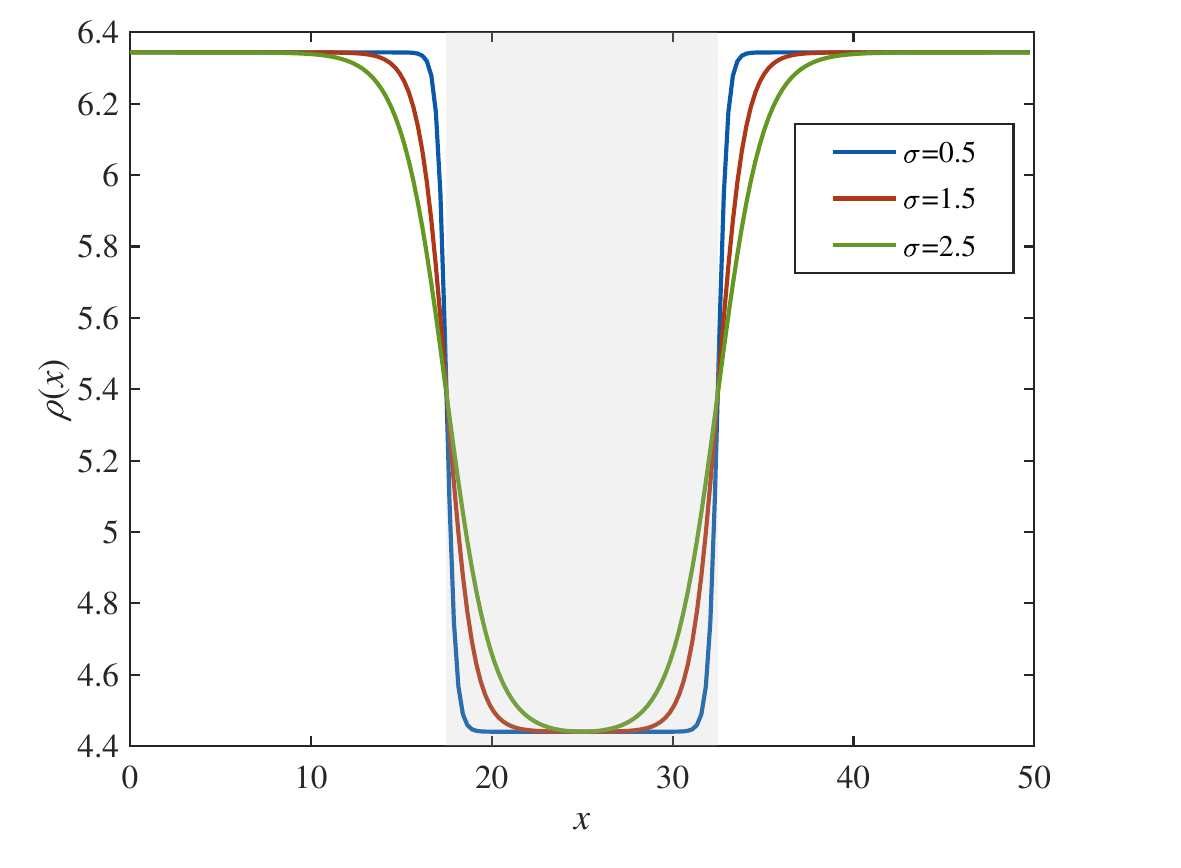}
\put(-220,190){(a)}~
\includegraphics[trim=1cm 7cm 0cm 8cm, clip=true, scale=0.44]{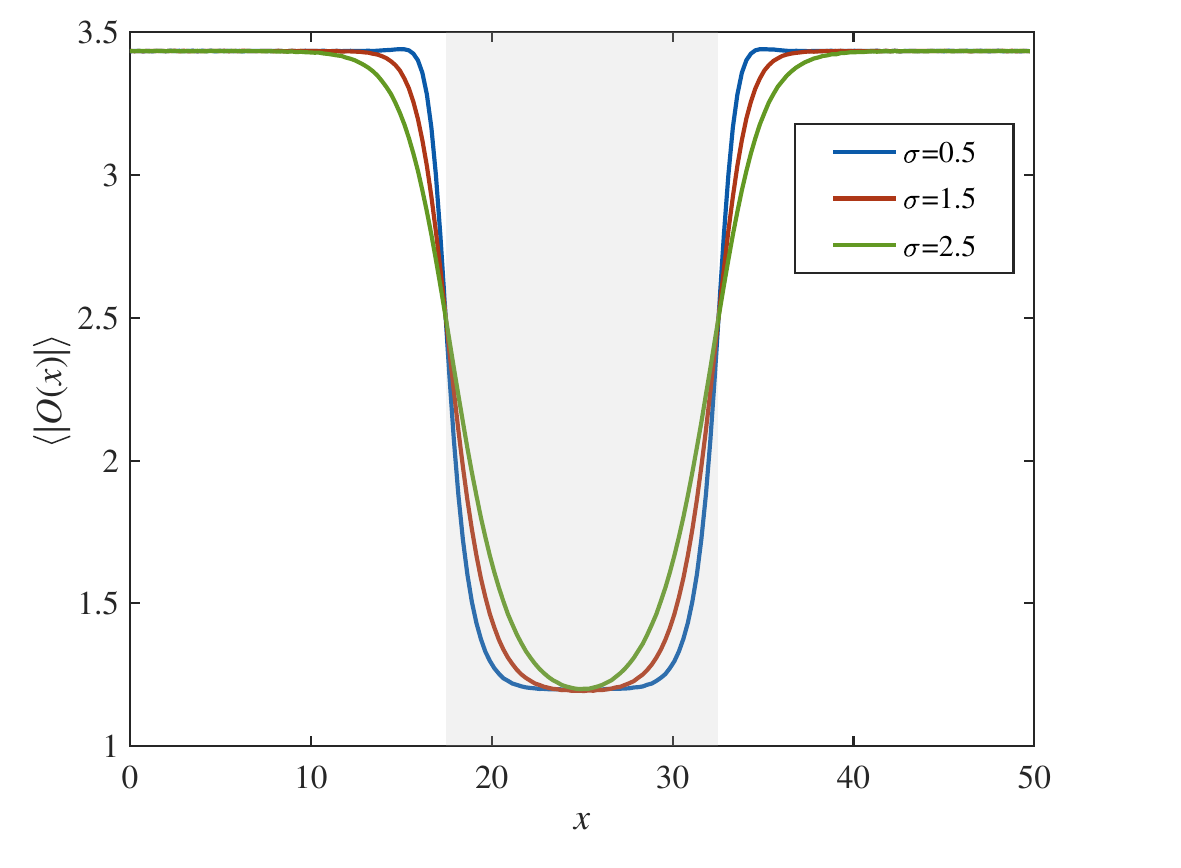}
\put(-220,190){(b)}~\\
\includegraphics[trim=1.cm 7cm 0cm 8cm, clip=true, scale=0.44]{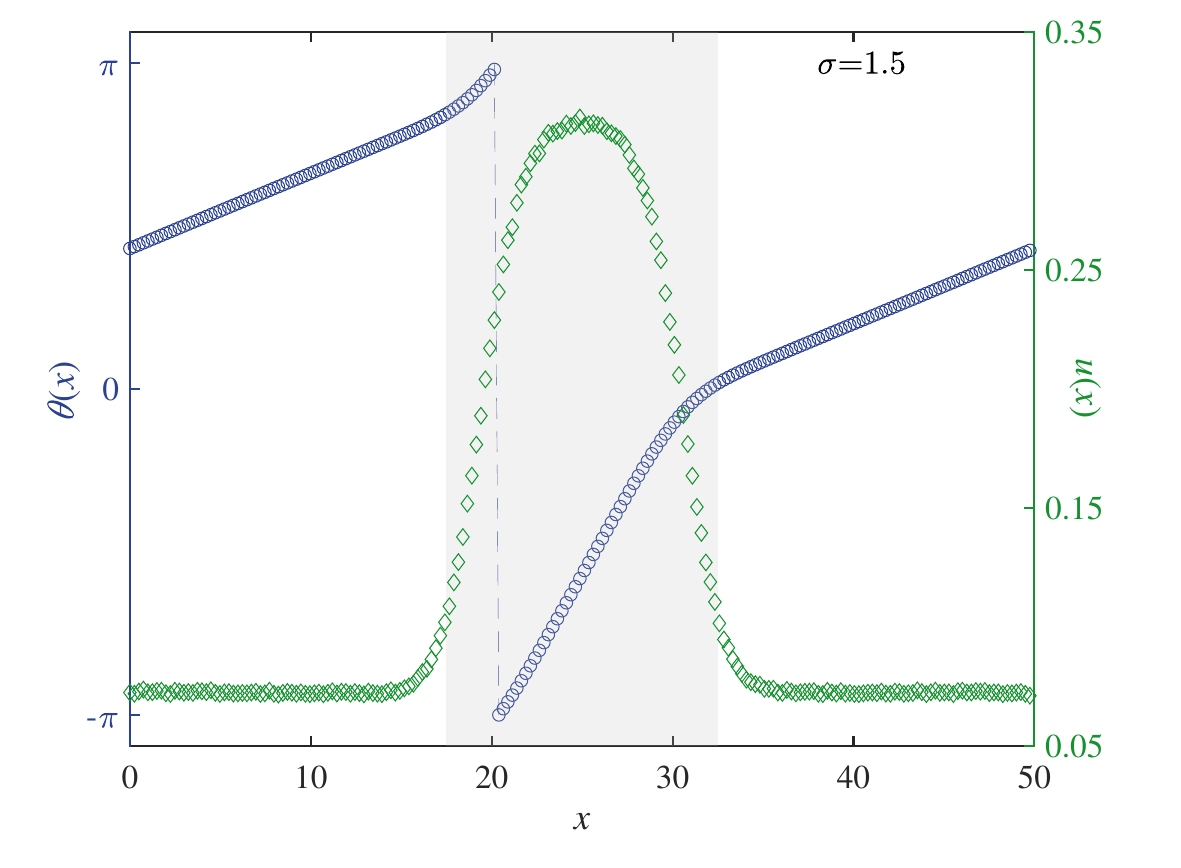}
\put(-220,190){(c)}~
\includegraphics[trim=1.cm 7cm 0cm 8cm, clip=true, scale=0.44]{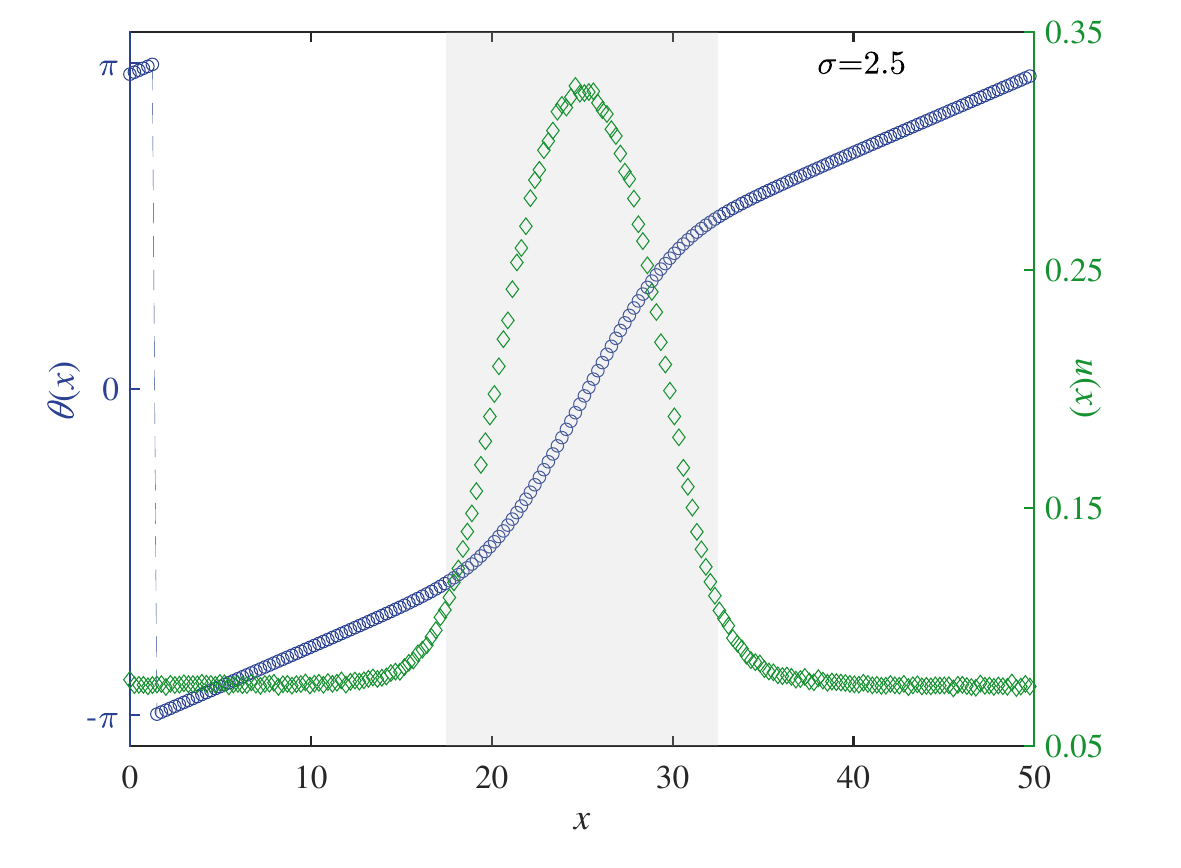}
\put(-220,190){(d)}~
\caption{The first row  shows the charge density $\rho(x)$ (panel (a)) and the condensate of the order parameter $\langle|O(x)|\rangle$ (panel (b)) for $\mathcal{L}=15$, $\epsilon=0.7$, $T_f=0.8T_c$ and different steepness \{$\sigma=0.5({\rm blue}),1.5({\rm red}),2.5({\rm green})$\}  in the final equilibrium state. The second row shows the configuration diagram of the phases $\theta(x)$ (blue circles)  and the corresponding  gauge-invariant velocity $u$ (green diamonds). The parameters of the panel (c)  are \{$\mathcal{L}=15$, $\epsilon=0.7$, $T_f=0.8T_c$, $\sigma=1.5$\} and the parameters of the panel (d) are \{$\mathcal{L}=15$, $\epsilon=0.7$, $T_f=0.8T_c$, $\sigma=2.5$\}. The shaded areas indicate junction regions.}\label{p6}
\end{figure}

\begin{table}[h]
\renewcommand{\arraystretch}{1.5}
\begin{center}
\begin{tabular}{|c|c|c|c|}
\hline
    $~~~~~~~~~$ & $~~~~u_1~~~~$ & $~~~~u_2~~~~$ & $(u_2-u_1) \frac{\mathcal{L}}{2}+\frac{L}{2} u_1$\\
    \hline
  $\sigma=0.5$ & $0.07$ & $0.2573$ & $3.1549$\\
    \hline
  $\sigma=1.5$ & $0.0729$ & $0.2555$ & $3.1917$\\
    \hline
  $\sigma=2.5$ & $0.0774$ & $0.246$ & $3.1995$\\
    \hline
\end{tabular}
\end{center}
\caption{Approximate values of $u_1$ and $u_2$ under various steepnesses of the Josephson junction for $\sigma=(0.5, 1.5, 2.5)$, and other parameters are fixed as  \{$\mathcal{L}=15$, $\epsilon=0.7$, $T_f=0.8T_c$\}. }
    \label{table2}
\end{table}%

{\bf {Case 2: Change steepness \{$\sigma=0.5, 1.5, 2.5$\} of the junction}}

Secondly, we study the effect of the steepness of the Josephson junction on the condensate of the order parameter as well as the gauge-invariant velocity. 

Figure \ref{p6}(a) exhibits the charge density configuration of Josephson junction at different steepnesses, with three different colors representing different values of steepness \{blue line ($\sigma=0.5$), red line ($\sigma=1.5$), green line ($\sigma=2.5$)\}.  From this plot we can see that the smaller $\sigma$ corresponds to steeper charge density. In Fig.\ref{p6}(b) we show the corresponding configurations of the condensate for various steepness. We see that the steeper the charge density is, the steeper the condensate is. We further compare the phase configurations (blue circles) and the corresponding gauge-invariant velocities (green diamonds) for different values of steepness in the second row of Fig.\ref{p6}, where Fig.\ref{p6}(c) and Fig.\ref{p6}(d)  have the steepness $\sigma=1.5$ and $\sigma=2.5$, respectively. The rest of the parameter values are the same \{$\mathcal{L}=15$, $\epsilon=0.7$, $T_f=0.8T_c$\}. For a more precise comparison of the effect of different steepnesses on the gauge-invariant velocities, we detail the velocities ($u_1$ and $u_2$) in both phases of the SSS Josephson junction in table \ref{table2}.  From table \ref{table2}, we observe that the steeper the charge density is (with smaller $\sigma$), the value of $(u_2-u_1) \frac{\mathcal{L}}{2}+\frac{L}{2} u_1$ is closer to $\pi$ as Eq.\eqref{eq:0} indicates.

\begin{figure}[h]
\centering
\includegraphics[trim=1cm 7.cm 0cm 8cm, clip=true, scale=0.44]{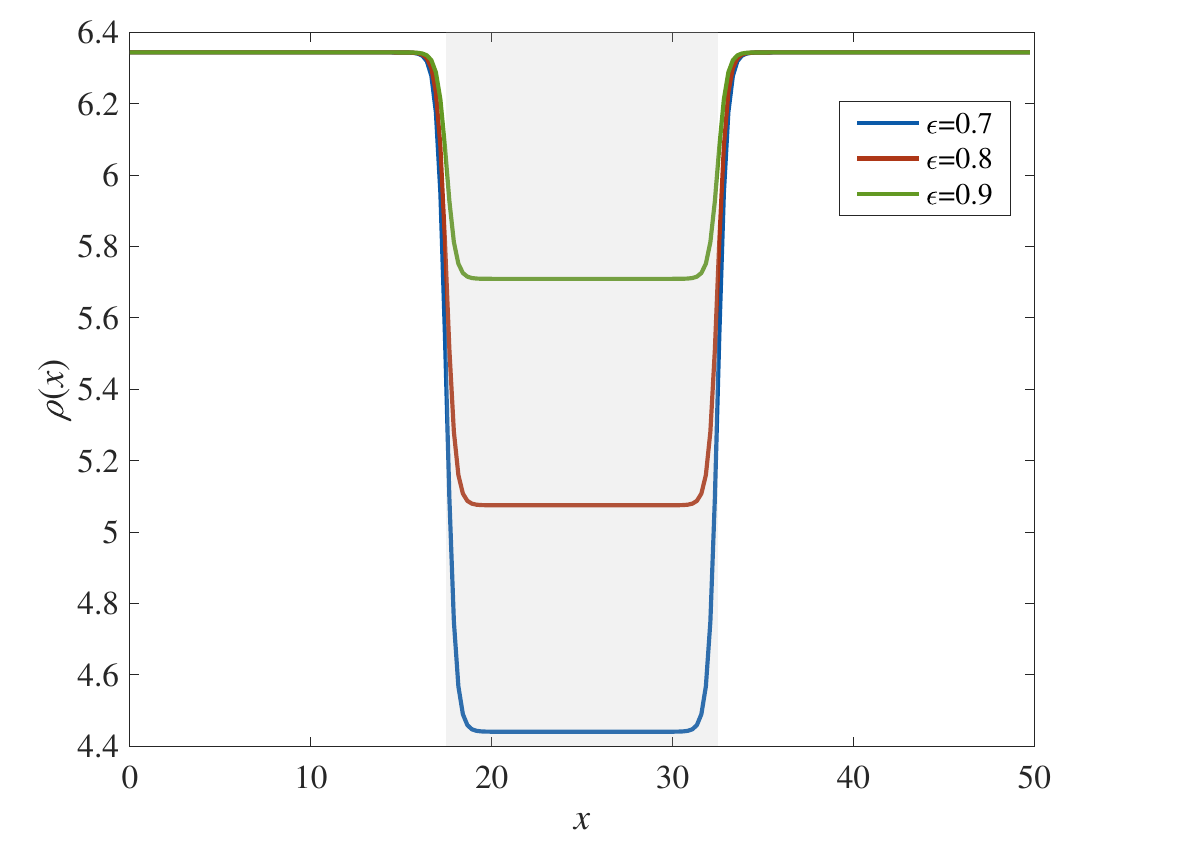}
\put(-220,190){(a)}~
\includegraphics[trim=1cm 7.cm 0cm 8cm, clip=true, scale=0.44]{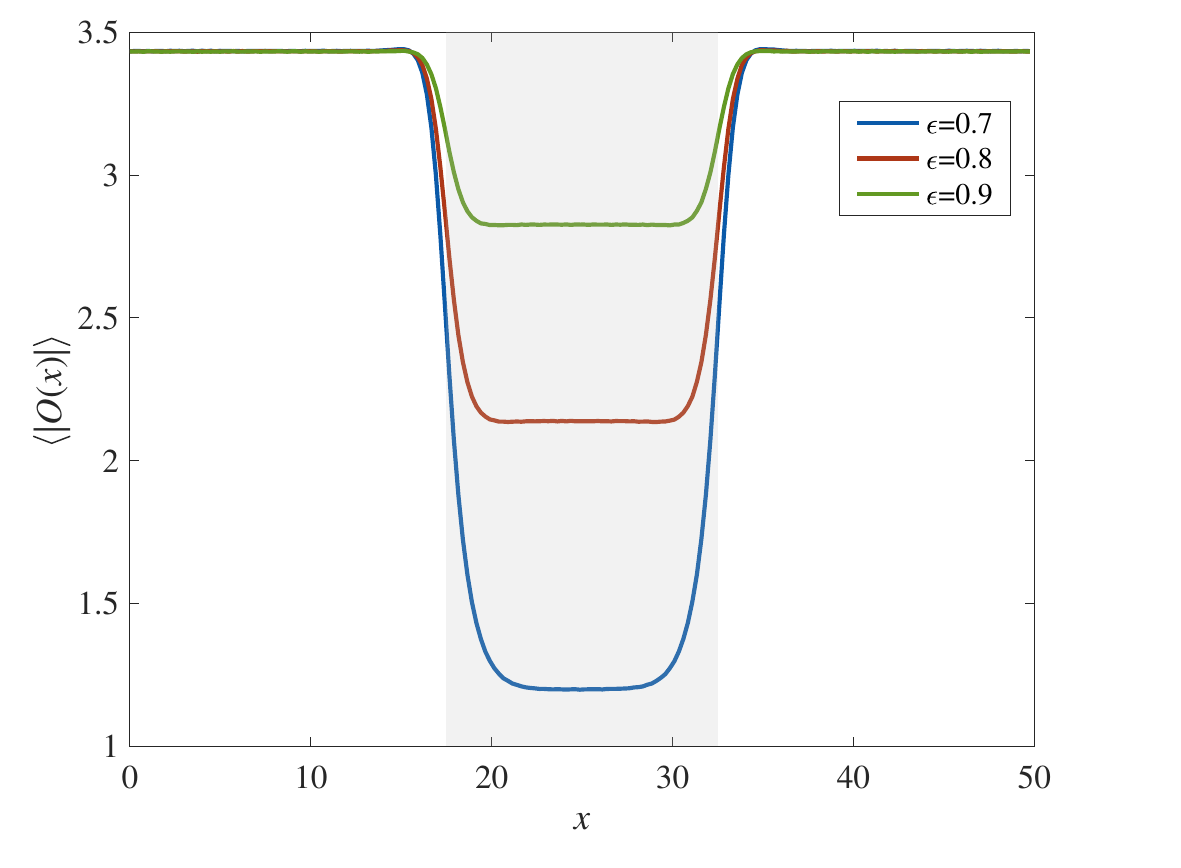}
\put(-220,190){(b)}~\\
\includegraphics[trim=1.cm 7.cm 0cm 8cm, clip=true, scale=0.44]{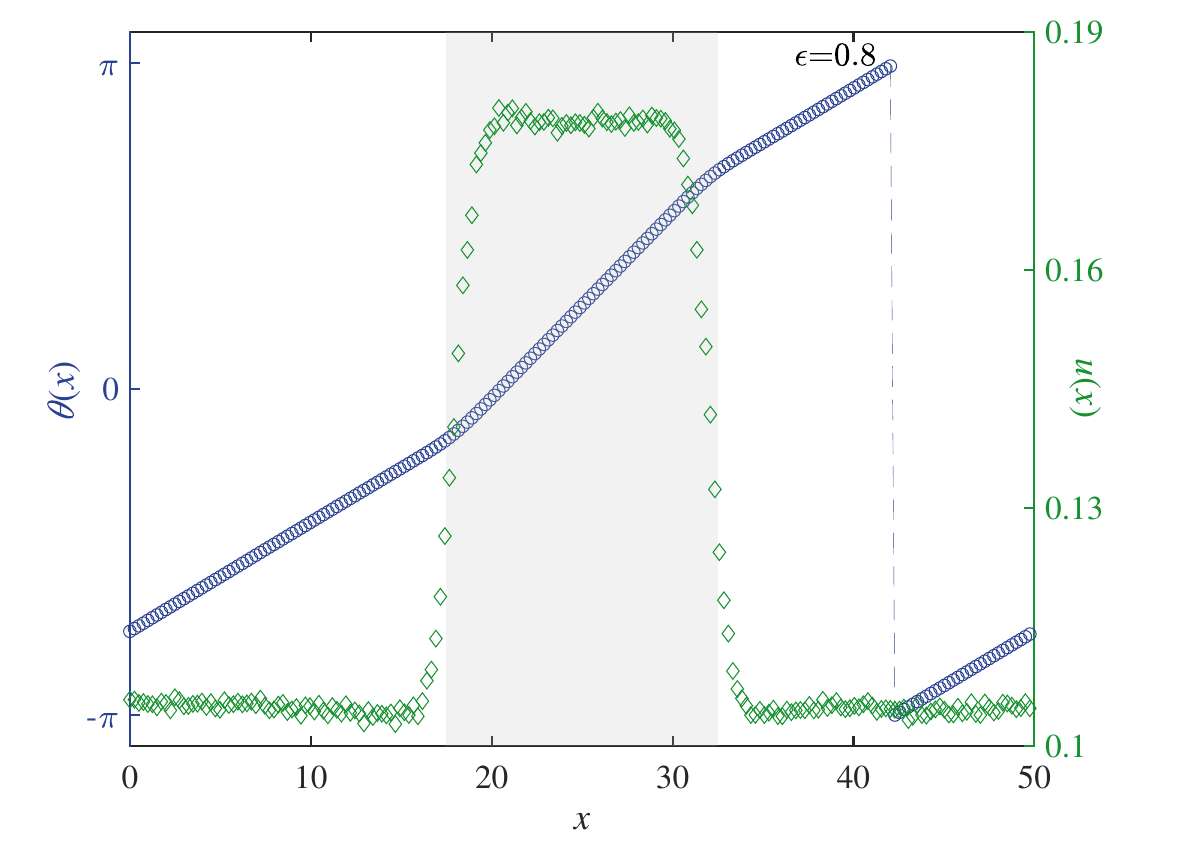}
\put(-220,190){(c)}~
\includegraphics[trim=1.cm 7.cm 0cm 8cm, clip=true, scale=0.44]{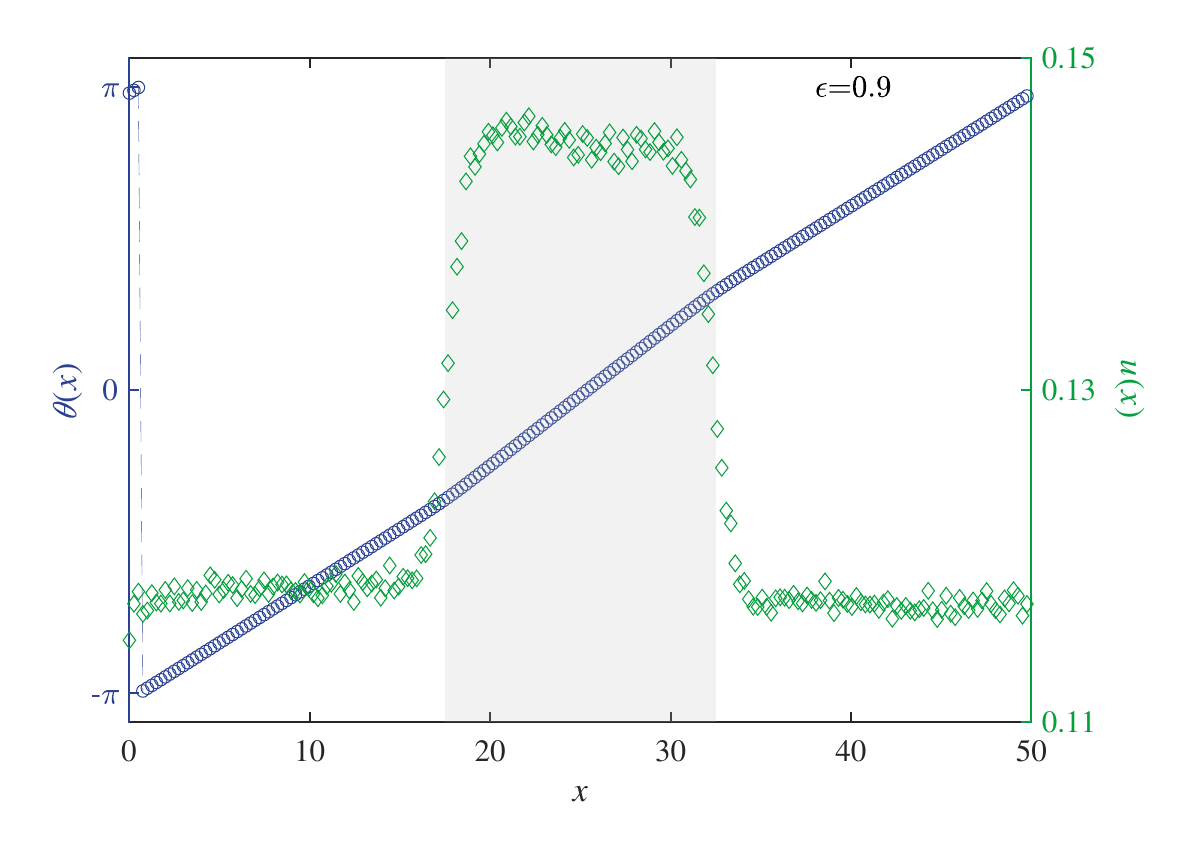}
\put(-220,190){(d)}~
\caption{The charge density $\rho(x)$ (panel (a)) and the condensate of the order parameter $\langle|O(x)|\rangle$ (panel (b)) for different depths \{$\epsilon=0.7({\rm blue}), 0.8({\rm red}), 0.9({\rm green})$\} at the final equilibrium state. Other parameters are $\mathcal{L}=15$, $\sigma=0.5$, $T_f=0.8T_c$;  The second row shows the configuration of the phases $\theta(x)$ (blue circles)  and the corresponding  gauge-invariant velocity $u$ (green diamonds). Other parameters for panel (c) are \{$\mathcal{L}=15$, $\sigma=0.5$, $T_f=0.8T_c$, $\epsilon=0.8$\} while for panel (d) are \{$\mathcal{L}=15$, $\sigma=0.5$, $T_f=0.8T_c$, $\epsilon=0.9$\} (right). The shaded areas indicate the junction regions. }\label{p7}
\end{figure}

\begin{table}[h]
\renewcommand{\arraystretch}{1.5}
\begin{center}
\begin{tabular}{|c|c|c|c|}
\hline
    $~~~~~~~~~$ & $~~~~u_1~~~~$ & $~~~~u_2~~~~$ & $(u_2-u_1) \frac{\mathcal{L}}{2}+\frac{L}{2} u_1$\\
    \hline
  $\epsilon=0.7$ & $0.07$ & $0.2573$ & $3.1549$\\
    \hline
  $\epsilon=0.8$ & $0.1049$ & $0.1753$ & $3.1501$\\
    \hline
  $\epsilon=0.9$ & $0.118$ & $0.1437$ & $3.1425$\\
    \hline
\end{tabular}
\end{center}
\caption{Approximate values of $u_1$ and $u_2$ under various depths of the Josephson junction for $\epsilon=(0.7, 0.8, 0.9)$, and other parameters are fixed as \{$\mathcal{L}=15$, $\sigma=0.5$, $T_f=0.8T_c$\}. }
    \label{table3}
\end{table}

{\bf{Case 3: Change depth  \{$\epsilon=0.7, 0.8, 0.9$\} of the junction}}

Thirdly, we will investigate the effect that the different depths of the junction bring to the system.

Figure \ref{p7}(a) exhibits the charge density configuration of Josephson junction at different depths, with three different colors representing different values of the depths \{blue line ($\epsilon=0.7$), red line ($\epsilon=0.8$), green line ($\epsilon=0.9$)\} from which we can see that the depth decreases with increasing $\epsilon$. The corresponding condensates are shown in Figure \ref{p7}(b), from which we can see that the depths of the condensates decrease as well with the increasing $\epsilon$. Fig.\ref{p7}(c) shows the configuration of the phases $\theta(x)$ (blue circles)  and the corresponding  gauge-invariant velocity $u$ (green diamonds)
with parameters \{$\mathcal{L}=15$, $\sigma=0.5$, $T_f=0.8T_c$, $\epsilon=0.8$\}, where the velocity  $u_1\approx 0.1049$ corresponding to the superconducting phase and $u_2 \approx 0.1753$ corresponding to the narrow superconducting phase, and the relationship satisfied between them is $(u_2-u_1) \frac{\mathcal{L}}{2}+\frac{L}{2} u_1=3.1501$, which is close to $\pi$. However, when $\epsilon=0.9$ (as Fig.\ref{p7}(d) shows) the velocity of superconducting phase $u_1\approx 0.118$ and the velocity of the narrow superconducting phase $u_2\approx 0.1437$, and $(u_2-u_1) \frac{\mathcal{L}}{2}+\frac{L}{2} u_1=3.1425$ which is much closer to $\pi$. The detailed data can be found in Table \ref{table3}. Therefore, we numerically verify that the relation Eq.\eqref{eq:0} is satisfied by the two velocities $u_1$ and $u_2$.

{\bf{Case 4: Change final temperature of quench \{$T_f=0.6T_c, 0.7T_c, 0.9T_c$\} }}

\begin{figure}[t]
\centering
\includegraphics[trim=1cm 7cm 0cm 8cm, clip=true, scale=0.44]{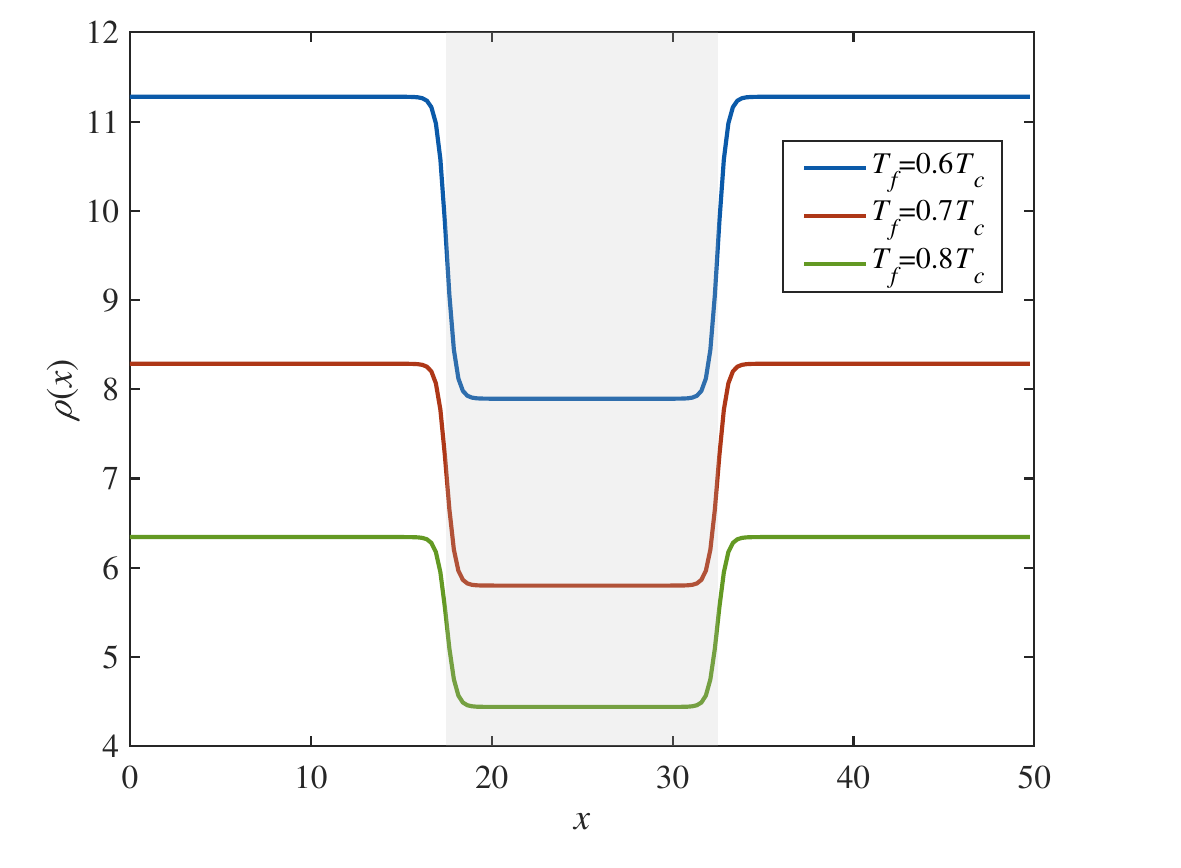}
\put(-220,190){(a)}~
\includegraphics[trim=1.cm 7cm 0cm 8cm, clip=true, scale=0.44]{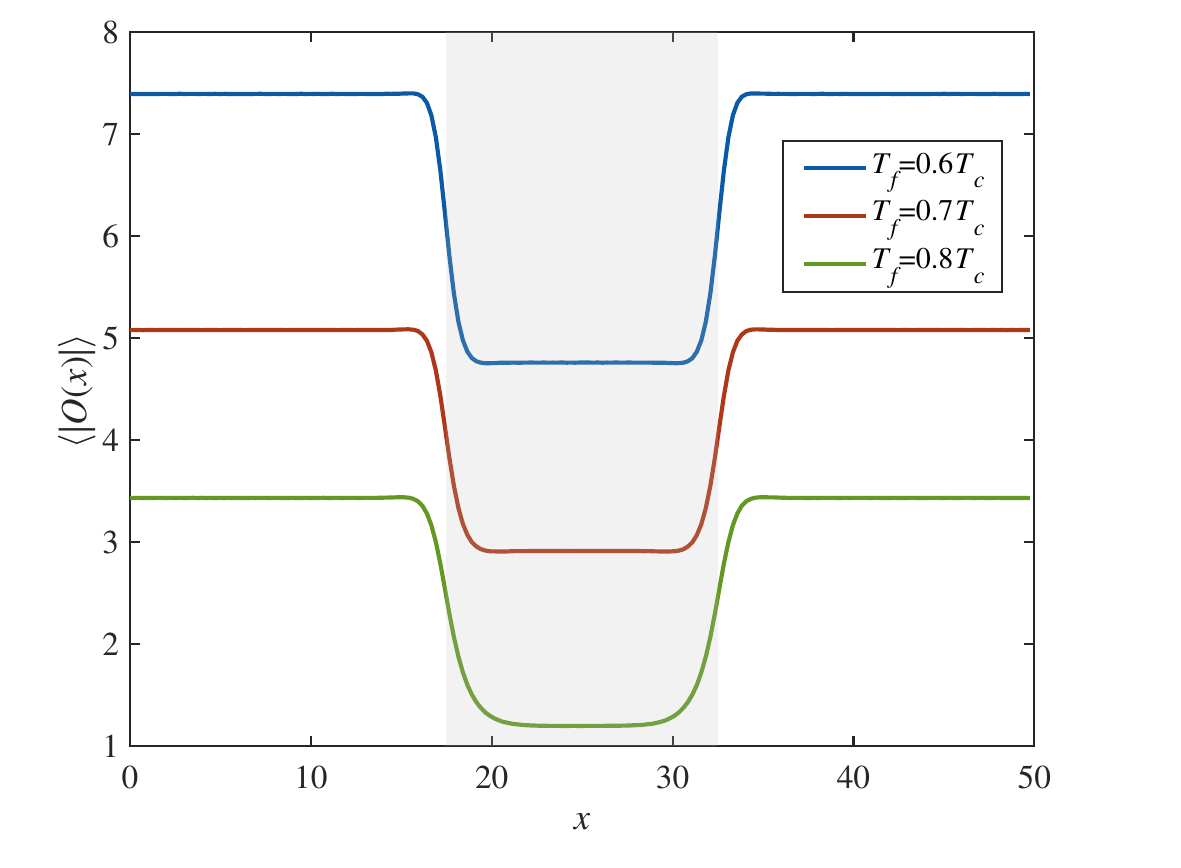}
\put(-220,190){(b)}~\\
\includegraphics[trim=1.cm 7cm 0cm 8cm, clip=true, scale=0.44]{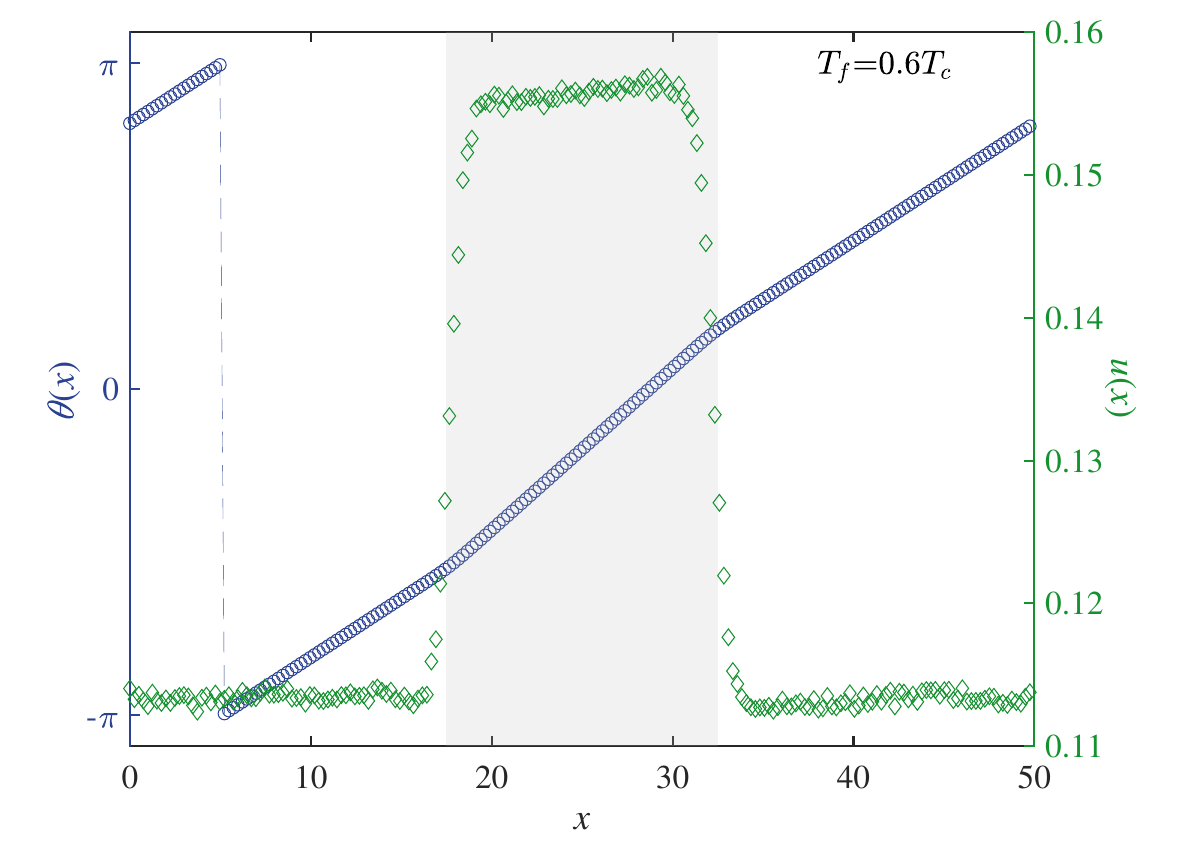}
\put(-220,190){(c)}~
\includegraphics[trim=1.cm 7.cm 0cm 8cm, clip=true, scale=0.44]{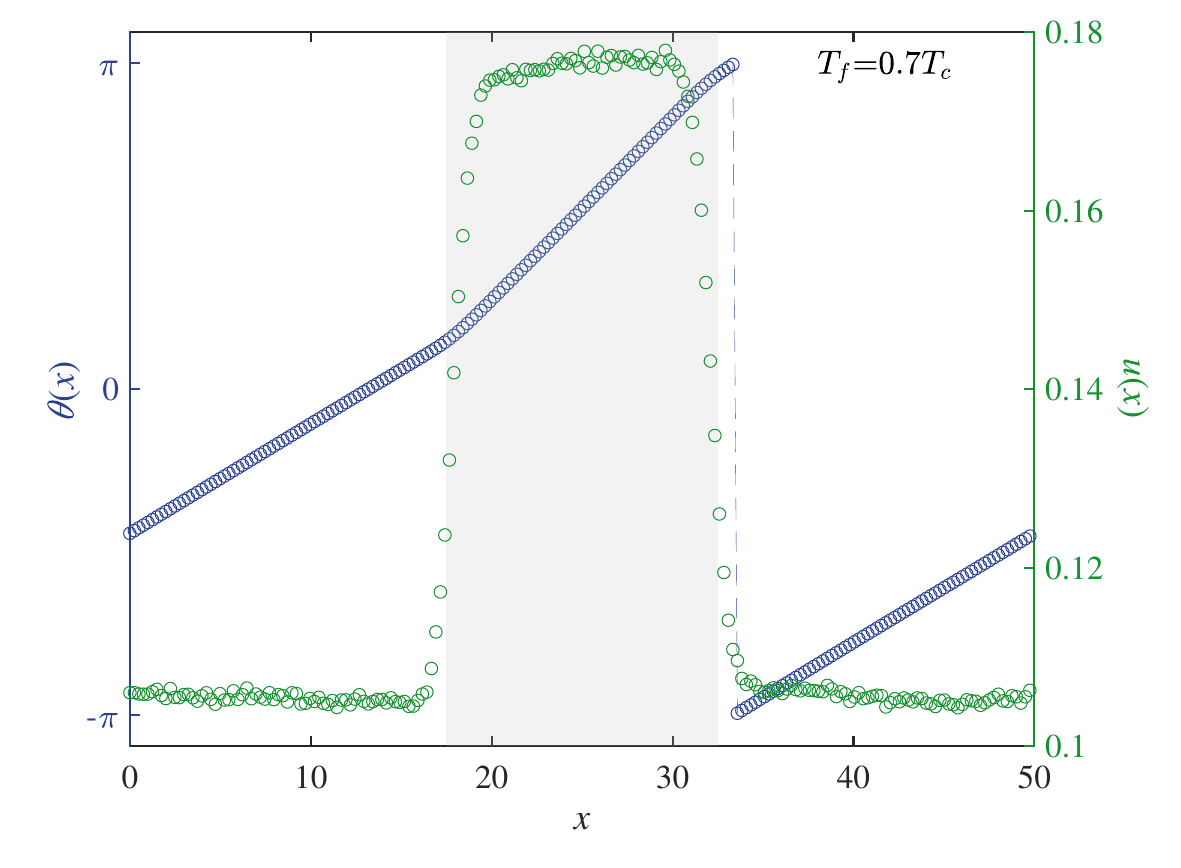}
\put(-220,190){(d)}~
\caption{The first row shows the charge density $\rho(x)$ (panel (a)) and the condensate of the order parameter $\langle|O(x)|\rangle$ (panel (b)) for different final temperatures of the quench in the final equilibrium state, where the blue, red and green lines represent final temperatures of $T_f $=\{$0.6T_c, 0.7T_c, 0.8T_c$\}. Other parameters are fixed as $\mathcal{L}=15$, $\epsilon=0.7$, $\sigma=0.5$. The second row  shows the configuration of the phases $\theta(x)$(blue circles)  and the corresponding  gauge-invariant velocity $u$ (green diamonds). The parameters of the panel (c) in the second row are \{$\mathcal{L}=15$, $\epsilon=0.7$, $\sigma=0.5$, $T_f=0.6T_c$\} while the parameters of the panel (d) are \{$\mathcal{L}=15$, $\epsilon=0.7$, $\sigma=0.5$, $T_f=0.7T_c$\}. The shaded areas indicate the junction regions.}\label{p8}
\end{figure}

The effect of the final temperature of the quench on the system is very interesting compared to the previous three cases. From Fig.\ref{p8}(a) and (b) we can find that the final temperature of the quench leads to changes in the charge density $\rho(x)$ and the condensate of the order parameter $\langle|O(x)|\rangle$, and the higher the final temperature of the quench is, the smaller the corresponding charge density and the condensate of the order parameter are. We further compare the phase configurations and the corresponding gauge-invariant velocities for different final temperatures of the quench in the second row of Fig.\ref{p8}, where the different parameters in Fig.\ref{p8}(c) and Fig.\ref{p8}(d) are $T_f $=\{$0.6T_c, 0.7T_c$\}, respectively.  The rest of the parameter values are fixed as \{$\mathcal{L}=15$, $\epsilon=0.7$, $\sigma=0.5$\}. For a more intuitive comparison of the effect of different final temperatures on the gauge-invariant velocities, we detail the velocities ($u_1$ and $u_2$) under both phases in the SSS Josephson junction in table \ref{table4}. From table \ref{table4}, it can be observed that lower final temperatures render the relation Eq.\eqref{eq:0} more precise.

 \begin{table}[h]
\renewcommand{\arraystretch}{1.5}
\begin{center}
\begin{tabular}{|c|c|c|c|}
\hline
    $~~~~~~~~~$ & $~~~~u_1~~~~$ & $~~~~u_2~~~~$ & $(u_2-u_1) \frac{\mathcal{L}}{2}+\frac{L}{2} u_1$\\
    \hline
  $T_f=0.6T_c$ & $0.1134$ & $0.1548$ & $3.146$\\
    \hline
  $T_f=0.7T_c$ & $0.1054$ & $0.1737$ & $3.147$\\
    \hline
  $T_f=0.8T_c$ & $0.07$ & $0.2573$ & $3.1549$\\
    \hline
\end{tabular}
\end{center}
\caption{Approximate values of $u_1$ and $u_2$ under various final temperatures of the quench for $T_f=(0.6, 0.7, 0.8)T_c$, and other parameters are fixed as  \{$\mathcal{L}=15$, $\sigma=0.5$, $\epsilon=0.7$\}. }
    \label{table4}
\end{table}%

After the discussion of the above four cases, we find such an interesting law: if the gauge-invariant velocity of the two phases of the SSS Josephson junction is closer to each other, the relationship between the two velocities Eq.\eqref{eq:0} will be more precise. When $u_1=u_2=u$, Eq.\eqref{eq:0} becomes $\frac{L}{2} u=\pi$, in which $u$ is constant at the final equilibrium state, indicating a persistent supercurrent along the ring. The integration of the velocity in the $x$-direction precisely equals the corresponding winding numbers (Eq.\eqref{eqw}) $W=\frac{1}{2\pi}\oint_\mathcal{C} u dx=\frac{L}{2\pi}u$ if $u$ is constant.  Both are self-consistent when $W = 1$.

\section{Discussion}\label{four}
By employing the KZM, we dynamically achieved the winding numbers of the order parameter in holographic superfluid ring with a weak link, i.e., a SSS Josephson junction. By varying the different parameters of the SSS Josephson junction (width, steepness, depth) and the final temperatures of the quench, we compared the configuration of the charge density and condensate of the order parameters,  we found that alterations in the width, steepness, and depth of Josephson junctions only affect the configuration of the charge density and condensate of the order parameters, but do not change the condensate values outside of the junction. However, differences in the final quenching temperature directly affect the values of the charge density and condensate of the order parameters. Furthermore, we conducted a comparison between the phase configurations of the order parameters and the gauge-invariant velocity at the final equilibrium state. We found that the phase $\theta$ finally became two `piecewise' straight lines in $x$-direction due to the presence of the Josephson junction, implying the existence of the supercurrent with two constant velocities. We further investigated the relationship between the gauge-invariant velocity in the superconducting state $u_1$ and the narrow superconducting state $u_2$ in the SSS Josephson junction and have discovered a relationship between the velocity of the two phases Eq.\eqref{eq:0}. By comparing the relationship between the two superfluid velocities, we observe that: increasing $\mathcal{L}$  (i.e. the wider the width of the junction) or $\epsilon$ (i.e. the shallower the depth of the junction) will make the numerical results more consistent with the Eq.\eqref{eq:0}; Similarly, decreasing $\sigma$ (i.e., the steeper the junction) or $T_f$ (i.e., the lower the final temperature) will bring the numerical results closer to Eq.\eqref{eq:0}.

\section*{Acknowledgements}

This work was partially supported by the National Natural Science Foundation of China (Grants No.12305067 and 12075143 ) and Shanxi Provincial Youth Scientific Research Project (Grants No.202303021222209 ). 

\end{document}